\documentclass[entropy,article,accept,moreauthors,pdftex,12pt,a4paper]{mdpi} 
\lastpage{2276}
\doinum{10.3390/e15062246}
\pubvolume{15}
\pubyear{2013}
\history{Received: 15 March 2013; in revised form: 21 May 2013 / Accepted: 30 May 2013 / \\
Published: 5 June 2013}

\usepackage{enumerate}
\usepackage{amsmath}
\def\tensor{\bar}
\usepackage{amssymb}
\usepackage{soul}
\usepackage{booktabs}

\Title{Bootstrap Methods for the Empirical Study of Decision-Making and Information Flows in Social Systems}

\Author{Simon DeDeo $^{1,}$*, Robert X.~D. Hawkins $^{1,2}$, Sara Klingenstein $^{1}$ and Tim Hitchcock $^{3}$}

\address{%
$^{1}$ Santa Fe Institute, 1399 Hyde Park Road, Santa Fe, NM 87501, USA; \linebreak E-Mails: hawkrobe@umail.iu.edu (R.X.D.H.); sara.klingenstein@gmail.com (S.K.)\\
$^{2}$ Cognitive Science Program, 819 Eigenmann, 1900 E. 10th St., Indiana University, Bloomington, IN 47406, USA\\
$^{3}$ Department of History, University of Hertfordshire, Hatfield, Hertfordshire, AL10 9AB, UK; \linebreak E-Mail: t.hitchcock@herts.ac.uk}

\corres{E-Mail: simon@santafe.edu.} 

\abstract{We characterize the statistical bootstrap for the estimation of information-theoretic quantities from data, with particular reference to its use in the study of large-scale social phenomena. Our methods allow one to preserve, approximately, the underlying axiomatic relationships of information theory---in particular, consistency under arbitrary coarse-graining---that motivate use of these quantities in the first place, while providing reliability comparable to the state of the art for Bayesian estimators. We show how information-theoretic quantities allow for rigorous empirical study of the decision-making capacities of rational agents, and the time-asymmetric flows of information in distributed systems. We provide illustrative examples by reference to ongoing collaborative work on the semantic structure of the British Criminal Court system and the conflict dynamics of the contemporary Afghanistan insurgency. }

\keyword{biological systems; cognition; social systems; information theory; statistical estimation; bootstrap; Bayesian estimation}

\begin{document}
\newpage
\section{Introduction} 

A major function of many biological and social systems is to encode, process, and share information. The functional forms of the information-theoretic quantities used to describe these aspects of a system are given to us by deduction from a remarkably small set of axioms. 

Estimation of these quantities is not trivial. When done carelessly, it can violate these underlying axioms, introduce spurious signals, lead to sensitive dependence on what should be innocuous choices of data representation, and create inconsistencies between estimation methods that otherwise should have been equivalent.

This paper addresses this problem. In particular, we first present a method---the statistical bootstrap---for estimating some of the most important information-theoretic quantities. The method preserves, approximately, the relevant axioms.

We shall show in particular that it outperforms both ``naive'' and Bayesian estimators in a regime of particular interest: when $n$, the number of samples, is at least as large as $k$, the number of bins, event-types, or categories. Not all empirical work satisfies this constraint, and much effort has been devoted to the under-sampled regime or to continuous data; however, a great many problems do, and these are the ones we are concerned with here. We shall show also that the bootstrap can provide reliable error estimates.

At the same time, we introduce, for the benefit of those working in the empirical sciences and who may be less familiar with the utility and power of information theory, the axioms and their direct utility in producing consistent and coherent accounts of the role that information, signaling, and prediction play in the real world. We do so by reference to two real-world examples, so as to provide an explicit guide for how information theory allows the phrasing, and answering, of vital questions.

We begin, in Section~\ref{cg}, with the entropy estimation problem. This section introduces the major technical themes of the paper: coarse-graining, the axiomatic foundations of information theory, and the use of the statistical bootstrap.

We then consider two uses of information theory in the real world. The first, considered in Section~\ref{distances}, is to measure, in a principled fashion, how much two distributions differ. We describe and interpret two measures, the well-known Kullback--Leibler divergence and the less well-known, but often \linebreak better-behaved, Jensen--Shannon divergence. We then show that quantifying the differences between distributions allows us to bound the probability of error made by participants in the system. We provide an illustrative example by reference to an on-going research project in the information-theoretic structure of the British Criminal Court system.

The second, considered in Section~\ref{mi-big}, is to measure the extent to which two patterns of behavior are synchronized. We emphasize the advantage of mutual information over less-principled measures such as the Pearson cross-correlation coefficient, with particular reference to the data processing inequality. We provide an illustrative example by reference to an on-going research project in the nature of \linebreak decision-making in the Afghanistan insurgency.

All of the results presented rely on the use of the statistical bootstrap. In Section~\ref{numerics} we detail numerical results on the use of this technique. We show how well the bootstrap corrects for bias, how well it preserves the relevant axioms (and their consequences), and how reliable its error estimates are. Our goal in this section is to provide an accurate guide for practitioners in the use of the bootstrap and to ground the explanations and accounts of the previous two sections. We conclude in Section~\ref{conc}.

\section{Estimating Entropy}
\label{cg}

Information Theory deals with probability distributions, $\vec{p}$, over outcomes. Its most fundamental quantity is entropy, which can be interpreted as the uncertainty of outcome when drawing from $\vec{p}$. Shannon's original paper~\cite{Shannon:1948p18105} establishes that the entropy function $H(\vec{p})$ takes a unique form, given the assumption of continuity and two additional conditions:
\begin{enumerate}[1.]
\item Uncertainty principle. When all $k$ entries of $\vec{p}$ are equal, $H(\vec{p})$ should be a monotonic, increasing function of $k$. \label{mono}
\vspace{-12pt}
\item Consistency under coarse-graining. $H(p_1,p_2,p_3)$ is equal to $H(p_1,p_2+p_3)+(p_2+p_3)H(p_2,p_3)$. \label{con1}
\end{enumerate}

Condition~\ref{mono} says that the uncertainty should rise when there are more possibilities (and all outcomes are equally likely). Condition~\ref{con1} says that if two outcomes are grouped, then the uncertainty is the uncertainty of the more coarse-grained description, plus (weighted) the uncertainty of outcomes from the grouped category.

Condition~\ref{con1} is not only the central axiom that leads to a unique mathematical form for the previously qualitative notion of uncertainty. Its recursive nature allows one to tie together descriptions of observed phenomena at vastly different resolutions, or coarse-grainings. The analysis presented in Section~\ref{distances}, for example, coarse-grains natural language texts to 116-dimensional feature vectors; however, other coarse-grainings, at 1031-dimensions and 26,740-dimensions in our analysis, are possible, and estimation methods that preserve Condition~\ref{con1} allow us to ask questions simultaneously about how much information is to be found at each level of description, and how much is lost when going from one description to another.

Our goal in this section will thus be the extension of Shannon's theory to discrete counts, $\vec{n}$, of observations such that Condition~\ref{con1} can, in some more or less exact fashion, be preserved. Depending on what limits the axioms are satisfied in (e.g., whether they hold only on average, or for large data), we consider functions to be more or less strict. The strictest demand requires 
\begin{enumerate}[$1^\prime$.]
\item Uncertainty principle. When all $k$ entries of $\vec{n}$ are equal, $\hat{H}(\vec{n})$ should be a monotonic, increasing function of $k$.\label{emp1}
\vspace{-12pt}
\item Consistency under coarse-graining. $\hat{H}(n_1,n_2,n_3)$ is equal to $\hat{H}(n_1,n_2+n_3)+\frac{n_2+n_3}{n}\hat{H}(n_2,n_3)$, where $n$ is the total number of observations.\label{emp2}
\vspace{-12pt}
\item Asymptotic convergence. As $n$ goes to infinity, $\hat{H}(\vec{n})\rightarrow H(\vec{p})$.\label{emp3}
\end{enumerate}

The simplest solution to this problem is to use the so-called naive estimator of the entropy,
\begin{equation}
\hat{H}(\vec{n}) = H\left(\left\{\frac{n_1}{n},\ldots,\frac{n_k}{n}\right\}\right)=H(\hat{p}(\vec{n}))
\label{naive}
\end{equation}
where $\hat{p}(\vec{n})$ is often called the empirical distribution. This satisfies Conditions~\ref{emp1} and~\ref{emp2} by construction, while satisfaction of Condition~\ref{emp3} is a consequence of the Asymptotic Equipartition Property. By a slight abuse of notation, we will write $H(\vec{n})$ in place of $H(\hat{p}(\vec{n}))$.

As proven in~\cite{Paninski:2003ff}, any estimator of entropy is necessarily biased---meaning that estimates of $H$ made on a finite sample will, on average, disagree with the asymptotic value. It is well known, for example, that the naive estimator above tends to underestimate the entropy of a system, and can be quite biased indeed for small $n$ (the first order correction), or small $p_i\ll 1/k^2$ (the second order correction). 

In an attempt to reduce the bias on the naive estimator, we can attempt a bootstrap correction.
\begin{equation}
H_\mathrm{corr}(\vec{n})=H(\vec{n})-\left[\langle H(\vec{n}^*) \rangle_{P(\vec{n}^* | \vec{n})}-H(\vec{n})\right]
\label{hcorr}
\end{equation}
where $\vec{n}^*$ is constrained to sum to $n$; here $P(\vec{n}^*|\vec{n})$ is the probability of drawing a set of $\sum n_i$ observations, $\vec{n}^*$, from an empirical distribution, $\hat{p}$, given by $\vec{n}$. Equation~(\ref{hcorr}) estimates the bias of $H(\vec{n})$ compared with $H(p)$ by estimating the average distance of $H(\vec{n}^*)$, where $\vec{n}^*$ is given by i.i.d. draws from $\hat{p}$, from$H(\vec{n})$. If the relevant properties of $\vec{p}$ are captured by $\hat{p}$, this should be a reasonable approximation.

\begin{figure}[H]
\centering
\includegraphics[width=5in]{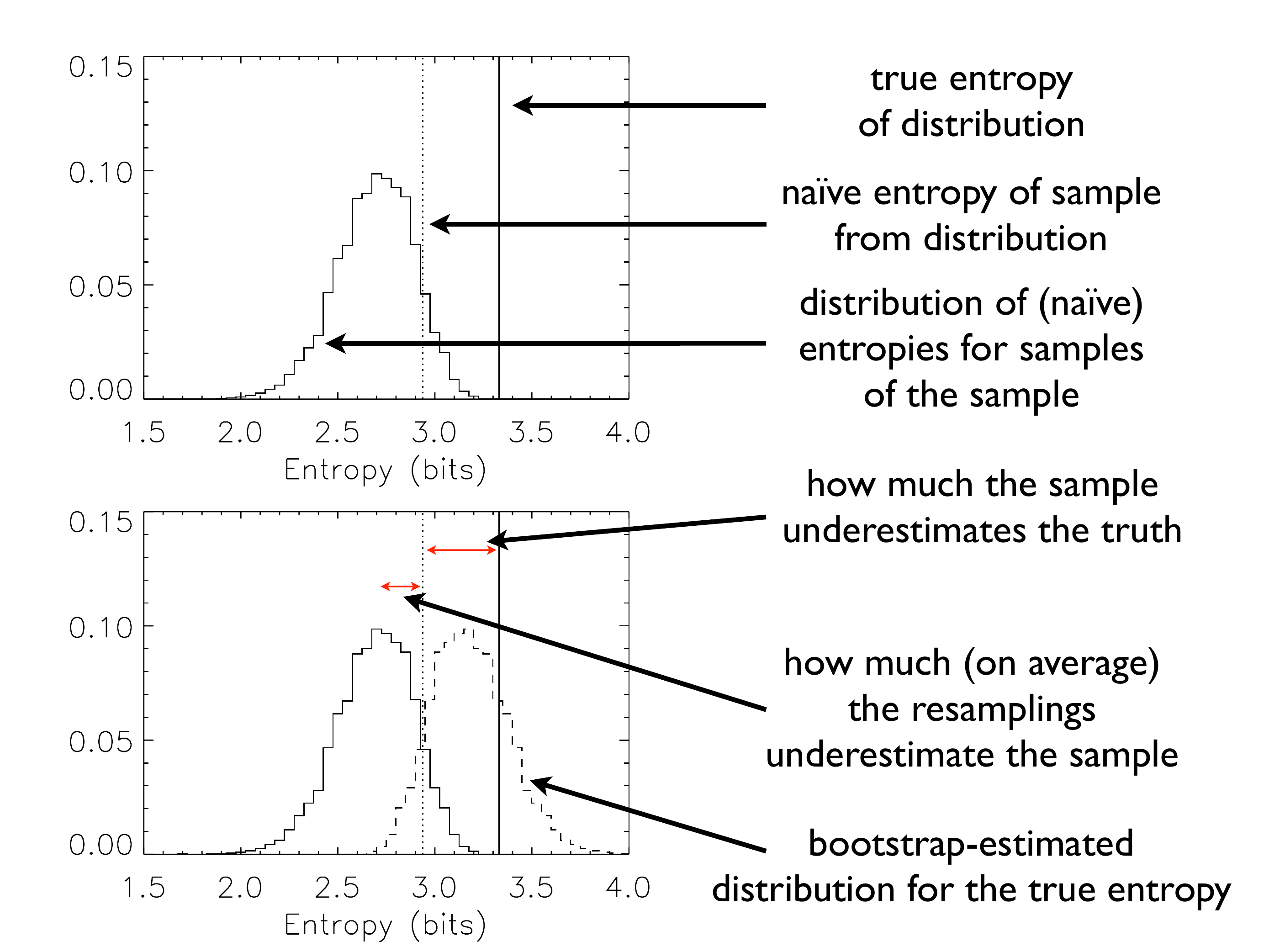}
\caption{A pedagogical example of the bootstrap in action.}
\label{bootstrap}
\end{figure}

Figure~\ref{bootstrap} illustrates explicitly the logic of the bootstrap, the implementation of Equation~(\ref{hcorr}), and the means by which one can obtain not only a bootstrap-corrected estimate but also error ranges for \linebreak that estimate.

The bias correction of Equation~(\ref{hcorr}) also violates the consistency conditions that should relate $H_\mathrm{corr}$ computed on fine- {\em vs.} coarse-grained distributions (Condition~\ref{emp2}). For example, we may expand the $\langle H(\vec{n}^*) \rangle$ term of Equation~(\ref{hcorr}) as
\begin{equation}
\langle H(\vec{n}^*) \rangle_{P(\vec{n}^* | \vec{n})} = \left\langle H(\vec{a}^*) + \frac{n^*_2+n^*_3}{n} H(\vec{b}^*) \right\rangle_{P(\vec{n}^* | \vec{n})}
\label{dir1}
\end{equation}
where $\vec{a}^*$ is $\{n_1^*,n^*_2+n^*_3\}$, and $\vec{b}^*$ is $\{n^*_2, n^*_3\}$. Conversely, if we had computed $\hat{H}_\mathrm{corr}$ in two steps, on the coarse-grained $\vec{n}$ and then the fine-grained subspace, the expansion of the equivalent terms would have given us
\begin{eqnarray}
\langle H(\vec{n}^*) \rangle_{P(\vec{n}^* | \vec{n})} & = & \langle H(\vec{a}^*) \rangle_{P(\vec{a}^* | \{n_1,n_2+n_3\})} + \frac{n_2+n_3}{n}\langle H(\vec{b}^*) \rangle_{P(\vec{b}^* | \{n_2,n_3\})}
\label{dir2}
\end{eqnarray}

Equations~(\ref{dir1}) and~(\ref{dir2}) are identical except for the fact that the second expectation value in Equation~(\ref{dir2}) fixes the prefactor to the maximum-likelihood value, while the expectation value in Equation~(\ref{dir1}) places no such constraint. Any coarse-graining enforced \emph{post hoc} will enforce a condition on the subset that is not enforced, by the bootstrap, in the aggregate. Consistency violations, however, are slight, and are at least a factor of ten less than other estimators in the literature. As we shall show in detail in Section~\ref{numerics}, the {\sc rms} error, even for $n$ equal to $k$, can be made quite small: for example, roughly one-hundredth of a bit, or 0.6\% of the average entropy value, for a sixteen-state system.

We turn now to the properties of information-theoretic quantities for the study of two central problems: distinguishability (Section~\ref{distances}) and synchronization (or signalling; Section~\ref{mi}).

\section{Distances between Distributions}
\label{distances}

In the year 1820, John Long was brought before a judge at the Old Bailey---the central criminal court of London, England---on a charge of breaking the peace. The full transcript of the trial was reported in the court proceedings of 18 September 1820 (see Appendix, Section~\ref{jlong}). Despite the seriousness of the charge (in the prior twenty years, seven in ten guilty verdicts for breaking the peace resulted in a death sentence) the full transcript for Long's trial is just under 400 words .

As part of an on-going collaborative research program on the nature of institutional \linebreak decision-making~\cite{sara}, we would like to know how much information the transcript contains about the outcome. Such a question is naturally phrased in terms of a distance or divergence: how ``far apart'' are trials, for example, that lead to guilty {\em vs.} not-guilty verdicts.

A contemporary of Long's, hearing of his being brought to trial, would have had quite a bit of information about what was likely to happen. She might know, as we do now, for example, that roughly three-quarters of all defendants at the Old Bailey the year before were given guilty verdicts (but only one in seven, when restricted to those indicted for breaking the peace). She might know more: that men were slightly more likely to be found guilty than women, for example (a difference of three percentage points in conviction rates).

Once the trial had begun an observer would expect to refine her beliefs about what would happen. Informally, we say that the transcript carries information about the outcome, and a good observer would be sensitive to it.

The transcript carries information, of course, about a great many things, including the legal and moral intuitions of the participants, their relative social status, and more or less reliable information about the actual events that the defendant is accused of taking part in. What we are interested in is the extent to which this talk, truthful or not, provides any signal of the underlying decision-making in the legal system itself. This question is independent of causal mechanisms: whether the transcript records an \emph{input} to the decision-making process, or whether it is simply a \emph{symptom} of hidden variables whose values are set by other means.

Answering this question will tell us a great deal: not only about the extent to which the \linebreak goings-on in the courtroom reflect actual features of the decision-making process, but also about the amount of information in principle available to actual observers---including participants who might alter, for strategic purposes, the information content and capacity of their behavior, or who might respond to information contained in the behavior of others.

To answer this question, we measure, using the data to hand, two distributions over transcript features. One distribution is constructed from trials with guilty outcomes, and one from those with not-guilty outcomes. As John Long's trial proceeds, we build up an empirical distribution over categories. If the transcript features we have identified are indeed information-bearing, in doing so we will learn something about which of the two distributions the trial is more likely to have been drawn from.

We focus on the lexical structure of the transcripts. We first measure how many times different words appear, dropping all information about the ordering of those words within a transcript. We draw on computational linguistic tools to split words by part of speech: for example, we distinguish whether ``dog'' is being used as a noun or verb. 

We then map these word counts to a more coarse-grained set of (one hundred and sixteen) semantic categories, and use this to build up an empirical distribution for the trial at hand. This amounts to an assumption of feature-independence: the claim that, while many aspects of language are clearly \linebreak order-dependent (syntax, for example, or the turns in which people speak), we shall consider only features whose arrival order does not itself carry information~\cite{word-order}.

Explicitly, then, we can then define two (empirical) distributions, $\hat{p}$ (for trials that have guilty \linebreak verdict outcomes),
\begin{equation}
\hat{p}_k = \frac{1}{n}\sum_{i=1}^{n_g} n_{ik}
\end{equation}
where $n_{ik}$ is the number of counts of words in semantic category $k$ found for (guilty-verdict) trial $i$, $n_g$ is the total number of guilty trials, and $n$ is the total number of semantic hits (\emph{i.e.}, the sum over all $n_{ik}$). We define $\hat{q}$ for trials with not-guilty outcomes similarly.

These approximations are sufficient to turn a qualitative question about the availability of information in trial transcripts into a form amenable to an information theoretic analysis. 

We consider two distinct ways to answer the transcript information question. Both are formulated in terms of a distance, difference or ``divergence'' between the distribution over categories for \linebreak guilty-outcome transcripts {\em vs.} the distribution over categories for the not-guilty outcomes.

Both methods seem to quantify essential aspects of the question. The first we consider, the \linebreak \emph{Kullback--Leibler divergence}, is well-known, but potentially unsuited to empirical work. The second, the \emph{Jensen--Shannon divergence}, is sufficiently well-behaved that it allows for bootstrap bias-correction and error estimation. 

In the final subsection, we show how the Jensen--Shannon divergence can be interpreted in terms of how well an optimal decision-maker can perform in gaining knowledge about the system in question, and present the \emph{Bhattacharyya bound}, which provides a strong bound on how well a rational observer can perform as more data comes in.

\newpage

\subsection{Kullback--Leibler Divergence}
\label{kld}

The first, Kullback--Leibler divergence, is an asymmetric measure. It can be interpreted as the answer to the following question: if the true underlying distribution is $p$, what is the asymptotic rate at which evidence accumulates against the alternative $q$?

We can derive the Kullback--Leibler divergence explicitly. We first write the probability of seeing a particular empirical distribution $\vec{n}$ given $\vec{p}$ as
\begin{equation}
P(\vec{n}|\vec{p}) = N\prod_{i=1}^k p_k^{n_k}
\end{equation}
and similarly for the distribution $q$; $N$ is a combinatoric constant common to both $P$. We can then write the asymptotic, geometric average of the ratio as 
\begin{equation}
A=\lim_{n\rightarrow\infty}\left(\frac{P(\vec{n}|\vec{p})}{P(\vec{n}|\vec{q})}\right)^{1/n}=\exp{\left(\lim_{n\rightarrow\infty}\frac{1}{n}\log{\frac{P(\vec{n}|\vec{p})}{P(\vec{n}|\vec{q})}}\right)}
\label{geo}
\end{equation}
Put picturesquely for the case of the Old Bailey, imagine a courtroom observer who has already heard an enormous amount of a trial, and that this trial is drawing from the distribution $p$. Each new observation will tend to confirm her belief that the trial is, indeed, drawing from $p$ (and not $q$), and will change her estimate of the relative probability of outcome, $P(p)/P(q)$, by a factor $A$.

The Kullback--Leibler (KL) divergence, $D(p,q)$, is then defined as the average value (in $p$) of the logarithm of $A$: $\langle \log_2{A}\rangle_p$. It can be written succinctly as
\begin{equation}
D(p,q)=\sum_{i=1}^k p_i\log{\frac{p_i}{q_i}}
\end{equation}
If there is a non-zero probability in $p$ of hearing a term that is impossible (zero probability) in $q$---a ``magic word'' that can only be produced by $p$---then there is a non-zero chance that the next word the observer will hear will make $P(\vec{n}|\vec{q})$ zero and so the Kullback--Leibler divergence will be infinite. Furthermore, since there is a non-zero probability for a resampling from $p$ and $q$ to create such a magic word, it is impossible to use bootstrap methods to correct for potential bias or to estimate error bars~\cite{magic-word}.

These infinities are particularly troublesome in empirical work where sparse sampling can generate magic words that do not reflect truly deterministic signals in the underlying process. In the next section, we investigate an alternative method for estimating the differences between distributions, the \linebreak Jensen--Shannon divergence, which is better behaved.

\subsection{Jensen--Shannon Divergence}
\label{jsd}

In contrast to the KL divergence, the Jensen--Shannon (JS) divergence is symmetric. It can be interpreted as the answer to the following question. Assume that a sample is drawn from either $p$ or $q$; $p$ is chosen with probability $\alpha$. How much is my uncertainty about which of the two distributions was used reduced by this single draw?

Again, put picturesquely, imagine our observer walks into a trial at the Old Bailey at random. Given the disposition of the judico-social institution as a whole, she has some (let us say correct) belief about the probabilities of outcomes. These leave her more or less uncertain about the fate of this particular defendant. Now she hears a single word spoken in the room. How much is her uncertainty (\emph{i.e.}, the entropy of the two-category choice guilty \emph{vs.} not-guilty) reduced? The answer is the Jensen--Shannon divergence, $J_\alpha(p,q)$.

More formally, the JS is the mutual information between a draw from one of the two distributions, and $Z$, a binary variable indicating which of the two distributions the process actually chose to draw from. When logarithms are base-two, this means that the Jensen--Shannon divergence is always between zero and unity (one bit).

The JS divergence has the functional form~\cite{Lin:1991ug}
\begin{equation}
J_\alpha(p, q) = \alpha D(\vec{p}, \vec{m}) + \beta D(\vec{q},\vec{m})
\end{equation}
where $\vec{m}$ is defined as
\begin{equation}
\vec{m}=\alpha\vec{p} + \beta\vec{q}
\end{equation}
with $\alpha$ between zero and unity and $\beta$ equal to $1-\alpha$. An equivalent definition, from which some identities become easier to derive, is
\begin{equation}
J_\alpha(\vec{p}, \vec{q}) = H(\alpha\vec{p}+\beta\vec{q})~-\alpha H(\vec{p})~-\beta H(\vec{q})
\label{def_h}
\end{equation}

While less well-known than the Kullback--Leibler divergence, the Jensen--Shannon divergence is always finite, and so it is possible to attempt bootstrap bias correction and error estimation. When taking the square root, it also satisfies the triangle inequality (see~\cite{Nielsen:2010vk} and references therein) and so can even function as a metric. 

Just as for entropy, we have a coarse-graining consistency relationship for the JS divergence. In particular, for two distributions $\vec{p}$, $\{p_1,p_2,p_3\}$ and $\vec{q}$, $\{q_1,q_2,q_3\}$, we can define the probability of landing in the $\{2,3\}$ subspace, $p_S$, as
\begin{equation}
p_S=\alpha(p_2+p_3)+\beta(q_2+q_3)
\end{equation}
We then have
\begin{eqnarray}
J_\alpha(\{p_1,p_2,p_3\},\{q_1,q_2,q_3\}) & = & J_\alpha(\{p_1,p_2+p_3\},\{q_1,q_2+q_3\}) \nonumber \\ & & +\,p_SJ_{\alpha(p_2+p_3)/p_S}(\{p_2,p_3\},\{q_2,q_3\}) \label{jsdconsistency}
\end{eqnarray}
where we silently renormalize probabilities in the subspace. Again, the consistency relationship shows the nested structure of information-theoretic reasoning: the additional information provided by more fine-grained distinctions appears, in weighted form, in the second term.

For two empirical distributions, $\vec{n}$ and $\vec{m}$, we can define the naive estimator using Equation~(\ref{def_h}), and from there define the bootstrap correction,
\begin{equation}
J_{\mathrm{corr}}(\vec{n}, \vec{m}) = 2\hat{J}(\vec{n}, \vec{m}) - \langle \hat{J}(\vec{n}^\star, \vec{m}^\star) \rangle_{P(\vec{n}^\star, \vec{m}^\star|\vec{n}, \vec{m})}
\label{js-boot}
\end{equation}
Bias correction, and error estimation, properties of Equation~(\ref{js-boot}) may be characterized in a similar fashion to the more common entropy and mutual-information estimation cases shown in detail in Section~\ref{numerics}. As for the bootstrap-corrected entropy and mutual-information, the bootstrap-corrected $J_\alpha$ leads to violations of the consistency relation, Equation~(\ref{jsdconsistency}). Violations are slight, and of the same order as found for the entropy itself. 

\subsection{The Bhattacharyya Bound}
\label{b-bound-section}

We now consider how information theory can place rigorous bounds on the abilities of observers both inside and external to the system. We will consider, in particular, how to measure a maximally-rational observer's \emph{rate of error}---the fraction of the time we can expect her to be wrong---when inferring facts about the system. Rather than measuring her error rate precisely, we will show how to bound it from~above.

Bounding the rate, as opposed to knowing it directly, will not be so great a loss. In making our subject maximally-rational, we have at the same time obscured some features of the system available to real-world observers. A real participant would have access to more information and thus be able to outperform the one we can describe. Thus, regardless of how well we estimate the error rate for our fiducial subject, we will only ever have (more or less strict) upper bounds.

Consider, once more, our observer at the Old Bailey, who watches a randomly-chosen trial and is interested in determining its outcome. Let us take the probability of a guilty outcome to be $\alpha$; for simplicity, assume that $\alpha$ is greater than 0.5. If our observer knows this, and nothing else, her best (``Bayes optimal'' or, informally, rational~\cite{Hellman:1970fn,jaynes}) guess, ``guilty'', will be wrong with probability $(1-\alpha)$.

When the observer acquires new information, her probability of error will decrease. By a theorem of Lin's~\cite{Lin:1991ug}, we can bound this updated probability of error. For a single observation, the probability of error is bounded from above by
\begin{equation}
P_e \leq \frac{1}{2}\left(H(\alpha,\beta) - J_\alpha(p,q)\right)
\end{equation}
In many cases, $J_\alpha(p,q)$ may be very small, and the bound on $P_e$ may not differ significantly from the bound given by prior knowledge. Indeed, in many real-world situations, this is to be expected---it would be remarkable, for example, if an observer were able to glean significant information about the progress of a trial from hearing only a single word! Unfortunately, the generalization of $P_e$ to the case of multiple observations does \emph{not} have a simple formulation in terms of the Jensen--Shannon divergence.

A number of different ways exist to approximate bounds on $P_e(n)$, the probability of error after $n$ observations. A commonly used one is the Bhattacharyya bound,
\begin{equation}
P_e(n)\leq \sqrt{\alpha\beta} \rho^n
\label{b-bound}
\end{equation}
where $\rho$ is
\begin{equation}
\rho = \sum_{i} \sqrt{p_i q_i}
\end{equation}
The Bhattacharyya bound is an approximation to the stricter Chernoff bound~\cite{cover}; in practice, it is very close~\cite{Chen:1976jm,ito1972approximate}, and far less computationally intensive to measure. Other approximations exist~\cite{Hashlamoun:1994gp}, but the functional form of the Bhattacharyya bound makes it possible to extend the bound to \linebreak multiple observations.

To give an example of the use of the Bhattacharyya bound, we can consider the statistics of criminal trials in the years prior to 1820. In the twenty year period between 1800 and 1820, the probability of a guilty verdict was 76\%---and so the associated error rate, before any observations are made, is 24\%. The associated Bhattacharyya bound, $P_e(0)$, is 43\%---not particularly tight (\emph{i.e.}, not close to the true value of 24\%). Indeed, given the ease with which the exact rate can be computed, it is not of great interest.

The power of the bound quickly becomes apparent, however. A standard coarse-graining we use in our investigations of this system sorts words into 116 possible categories. For the twenty year period between 1800 and 1820, the $\rho$ term for this particular coarse-graining, splitting guilty {\em vs.}~not-guilty verdicts, is approximately $0.9980\pm0.0002$---very close to, but not exactly, unity \cite{bs-jsd}. With knowledge of $\rho$, we can then compute the probability of error given an \emph{arbitrary} number of words. The error rate drops to 5\% after approximately eight hundred words, and to 1\% at 1620.

The Bhattacharyya bound has an important limitation: it refers to prediction not upon sampling repeatedly from a particular instance of a class (a particular trial), but on sampling from the overall distribution. These two cases may, or may not, be equivalent. An obvious way for the equivalence to fail is for the underlying bag-of-words model to fail: if a trial's semantic features are not independent draws, but that prior text within the trail alters the distribution from which the remainder is drawn.

A less obvious way for the assumption to fail is if there are multiple sub-classes. It may, for example, be the case that a guilty-verdict trial can take two forms: independent draws from $p_1$, or independent draws from $p_2$. If the two sub-classes are equally likely, then we will measure $p$ to be the average of $p_1$ and $p_2$, but there is no such thing as a trial that draws from $p$~\cite{further-clarify}. Note that when the sub-class membership is unknown, the second failure mode also leads to conditional dependence---each draw gives better information about which sub-class has been chosen, and thus alters one's beliefs about \linebreak subsequent draws.

For single draws, this does not matter; it only becomes apparent when making multiple draws; if the $n$ draws are represented as $\{n_i\}$, and the two sub-classes are equally likely, we have
\begin{equation}
P(p_1~\mathrm{or}~p_2|\vec{n}) \propto \alpha\left[\left(\prod_{i=1}^n p_1(n_i)\right) +\left(\prod_{i=1}^n p_2(n_i) \right)\right]
\label{two-class}
\end{equation}
which is distinct from the single-class case used to derive the Bhattacharyya bound,
\begin{equation}
P(p|\vec{n}) \propto \alpha\prod_{i=1}^n \left(p_1(n_i)+p_2(n_i)\right)
\label{one-class}
\end{equation}
which contains cross-terms. The functional form of Equation~(\ref{two-class}) makes it impossible to derive an equally simple version of Equation~(\ref{b-bound}) for the multiple-class case.

\begin{figure}[H]
\centering
\includegraphics[width=4.5in]{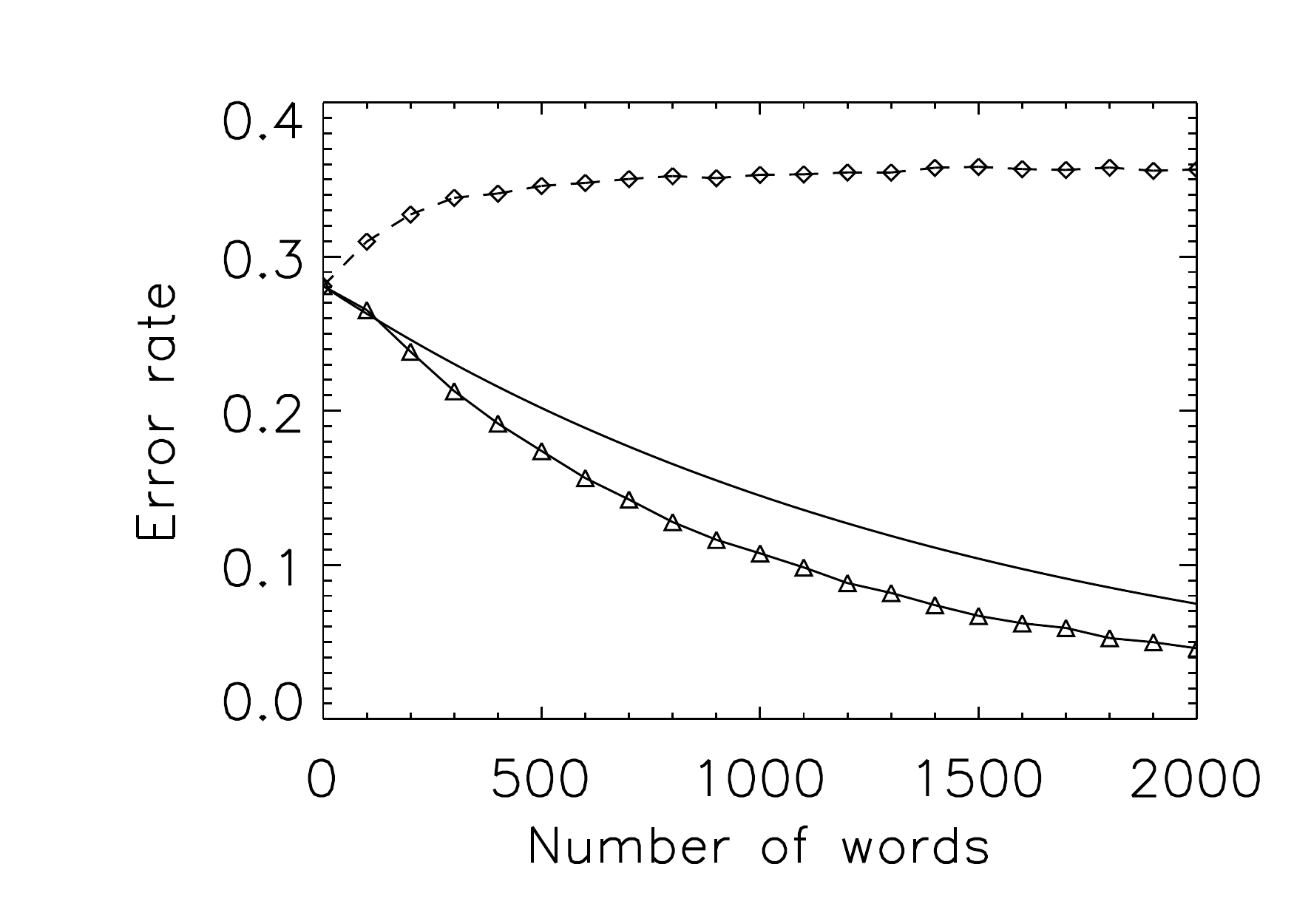}
\caption{Prediction error curves and the existence of multiple classes. Solid curve: the Bhattacharyya bound for prediction of trial outcome for the period 1800 to 1820. Triangle symbols and solid line: actual prediction error, when drawing samples (words) from all trials within a class (guilty or not-guilty). As expected, the curve lies strictly below the Bhattacharyya bound. Diamond symbols and dashed line: actual prediction error, when drawing samples from a single trial. The prediction error actually rises (more samples lead to a less accurate prediction), suggesting that the underlying model (trials sample from one of two distributions) is incorrect. We restrict the set of trials here to those with at least one hundred (semantically-associated) words, so as to make the resampling process \protect\linebreak more accurate.}
\label{b-bound-fig}
\end{figure}

We can study the validity of the assumptions of the Bhattacharyya bound by comparing the predicted bounds on the error rate with the actual success we have on predicting the outcomes of trials in the dataset itself. Figure~\ref{b-bound-fig} plots (1) the Bhattacharyya bound; (2) the error rate if each observation samples at random from the set of all trials with a particular outcome; and (3) the error rate if each observation samples only from a single, randomly chosen, trial (and predictions are made using Equation~(\ref{one-class})).

As expected, (2) is bounded by (1), but the error rate for (3) actually \emph{rises}: we do worse at predicting the verdict for a particular trial when given the transcript. Such an outcome strongly suggests, of course, that assumption of a single class is wrong; conversely, that there are different ways to be found guilty, and that these differences leave signatures in the semantic features of the trials themselves. It is a form of ergodicity breaking that one expects to be common in social systems: a single (guilty verdict) trial will not sample the full space of possible (guilty verdict) trial features. Correct estimation of the error rate for a real-world observer of a particular trial requires one to estimate the number of sub-classes within each verdict.

\subsection{Summary}

Measuring the extent to which two distributions differ is a common question in an \linebreak information-theoretic context. Differences between distributions are naturally interpreted as decision problems: how well ideal observers can distinguish different outcomes. Different measures have different interpretations, and while these differences appear subtle, their properties may make them more or less useful for empirical study.

The KL divergence can be interpreted in terms of an asymptotic rate at which information for a particular hypothesis is accumulated, given that the hypothesis is true. It has the unfortunate property of becoming infinite under conditions we would expect in the real world.

Meanwhile, the JS divergence is well behaved. It can be interpreted as the amount of information (in bits) relevant to distinguishing two outcomes that is contained in a single observation of the system.

Finally, the Bhattacharyya bound extends the Jensen--Shannon divergence to the case of multiple observations. Care must be taken in its use, since a binary decision task may involve multiple sub-classes; the bound is strictly true only when considering draws from the overall distribution.

\section{Correlation, Dependency and Mutual Information}
\label{mi-big}

On {3 April 2005}, in Spin Boldak, a town in Kandahar province, members of the Afghanistan insurgency remotely detonated a bomb concealed in a beverage vendor cart. The resulting explosion killed two people: a civilian and a police officer. The full report, as used in the analysis of this section, appears in Appendix Section~\ref{sigact}. It was one of eight events in the country recorded by members of the International Afghanistan Security Forces (ISAF) that day. To what extent was this event coordinated with others that day, week, or year?

The event described above is drawn from the Afghan War Diary (AWD) database, a remarkably detailed account of the Afghanistan conflict and the actions by both the insurgency and ISAF. The \linebreak open-source nature of the release has led to a number of efforts to characterize the data~\cite{larry-article,2012PNAS..10912414Z}. It is likely to become a standard set for both the analysis of human conflict and the study of empirical methods for the analysis of complex, multi-modal data. The release amounts to roughly $70,000$ SIGACTs (``Significant Activity Reports''), which record detailed information about individual events.

After filtering the data, we distinguish between SIGACTs that record insurgent-initiated events {\em vs.} those that record ISAF-initiated events~\cite{analysis-detail}. We then choose either of the two sets, group the events by the day on which they occurred, and generate symbolic time series for each province. 

We coarse-grain the complex information available in the SIGACT data by means of a four-state codebook, where codes are assigned based on the severity of violence. In order of increasing severity, Code 0 is when no events are recorded in the province that day; Code 1 is when events are recorded, but no injuries or deaths are associated; Code 2 is when one or two injuries or deaths are recorded; Code 3 is when more than two injuries or deaths are recorded. Taking the example of Kandahar province on April 3rd, there was, in addition to the Spin Boldak bombing, one other insurgent-initiated event recorded: a second IED (improvised explosive device) explosion with no reported injuries or deaths. Based on these two facts, we assign the insurgent time stream for Kandahar province Code ``2''.

Central to an understanding of modern insurgencies is measuring (1) the level of communication and coordination among insurgent groups~\cite{Sanin:2010cx}; and (2) signaling of intents and abilities, both among groups and between groups, government actors and the civilian population~\cite{GutierrezSanin:2008el,green}. This leads directly to the information theoretic question of the extent to which this event is coordinated with other events in the system, and the extent to which the other events may or may not have played a signaling role. 

As part of an ongoing collaborative investigation~\cite{hawkins} we would like to know---or at least bound---the minimal amount of information shared between systems for the purposes of synchronization or response, and how the temporal structure of this shared information changes in time. Examining the data within the axiomatic framework of information theory is likely to provide a novel approach to longstanding conceptual and quantitative questions at the center of the study of human conflict.

Because of its centrality to the study of decentralized insurgency, we focus in this section on the question of signaling and coordination. We focus in particular on two neighboring provinces (Kandahar and Helmand) in a single year (2007) to show the kind of questions information theory allows us to pose and the provocative answers it provides.

\newpage

\subsection{Mutual Information}
\label{mi}

Mutual information is a specialization of the divergence measures considered in the previous section. In particular, it measures the KL distance between two distributions: a joint distribution, $p_{ij}$, and one derived from $p_{ij}$ but in which the processes are forced to be independent,
\begin{equation}
I(p_{ij}) = D(p_{ij}, p_{i\cdot}p_{\cdot j})
\label{mi-def}
\end{equation}
where the marginals are
\begin{equation}
p_{i\cdot} = \sum_{j=1}^{k_j} p_{ij}
\end{equation}
and similarly for $p_{\cdot j}$. If the space of events labelled by $i$ is $A$, and $B$ for $j$, we often write $I(A,B)$ \linebreak for Equation~(\ref{mi-def}).

A standard, and useful, interpretation of mutual information is the average reduction in uncertainty (entropy) of the value of a sample from $A$, given knowledge of the value of a sample from $B$. This means, among other things, that the entropy of $A$ is an upper limit on the mutual information between $A$ and any other variable. 

As with entropy and the Jensen--Shannon divergence, the naive estimator can be used to define a bootstrap-corrected version,
\begin{equation}
I_\mathrm{corr}(\tensor{n}) = 2I(\tensor{n}) - \langle I(\tensor{n}^*) \rangle_{P(\tensor{n}^*|\tensor{n})}
\label{icorr}
\end{equation}
where coarse-graining consistency is now preserved only approximately.

By contrast with $H_\mathrm{corr}$, it is possible for $I_\mathrm{corr}$ to be less than zero: observations that happen to produce a precisely factorizable empirical distribution, for example, so that $I(\tensor{n})$ is zero, will have non-zero probability to resample to a distribution that \emph{does} produce correlations. Thus, the information inequality, $I(X;Y)\geq0$ with equality if and only if $X$ and $Y$ are independent, does not hold for $I_\mathrm{corr}$. This parallels the difficulty, in many empirical studies, of establishing complete conditional independence given finite data~\cite{Hutter:2001wt,Zaffalon:2002vx}, and requires reliable estimation of error ranges in order to prevent reporting of \linebreak nonexistent relationships.

Much like the quantities of the previous section, mutual information does not directly measure causation. This can be seen explicitly in the symmetric structure of Equation~(\ref{mi-def}), where $I(A,B)$ is equal to $I(B,A)$. There are a number of methods for finding (approximate) answers to causal questions~\cite{2011arXiv1102.1507W,2000PhRvL..85..461S}; a common starting point is to examine time-lagged mutual information: the methods we describe here, and the means by which they are characterized, are equally amenable to the case where $A$ is time-lagged relative to $B$. 

The mutual information for the two provinces, given only knowledge of four-state codes, is \linebreak $0.04\pm0.02$~bits (bootstrap corrected); same-day knowledge of the events gives a small, but detectable, boost to predictive ability, indicating some pathway for the sharing of information between the two processes. We emphasize that such a pathway may not be direct, and may involve common cause nodes that act not as a conduit of information but as a synchronizing signal. 

Obvious exogenous, common causes include external political factors such as a national election, and seasonal weather patterns that make it hard for the insurgency to act during harsh winters. For example, if the insurgency usually conducts daily high-severity events (codes 2 and 3) but a harsh winter makes this impossible, knowing the severity of an event in Kandahar will disclose the season \linebreak (winter/not-winter, one bit), and lead to (at least) one bit of mutual information between Kandahar and Helmand provinces---without any direct causal influence.

\begin{figure}[H]
\centering
\includegraphics[width=4in]{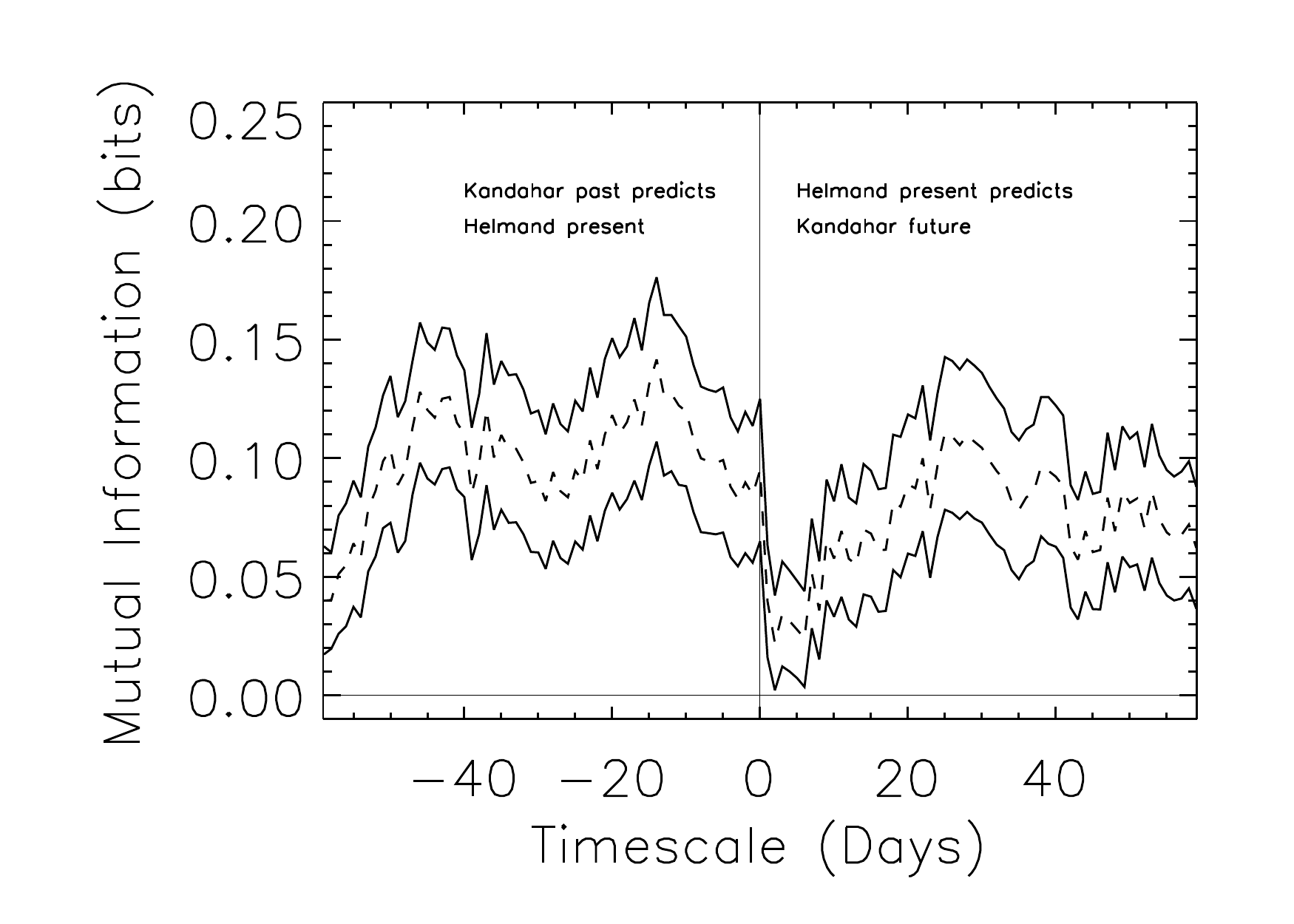}
\includegraphics[width=4in]{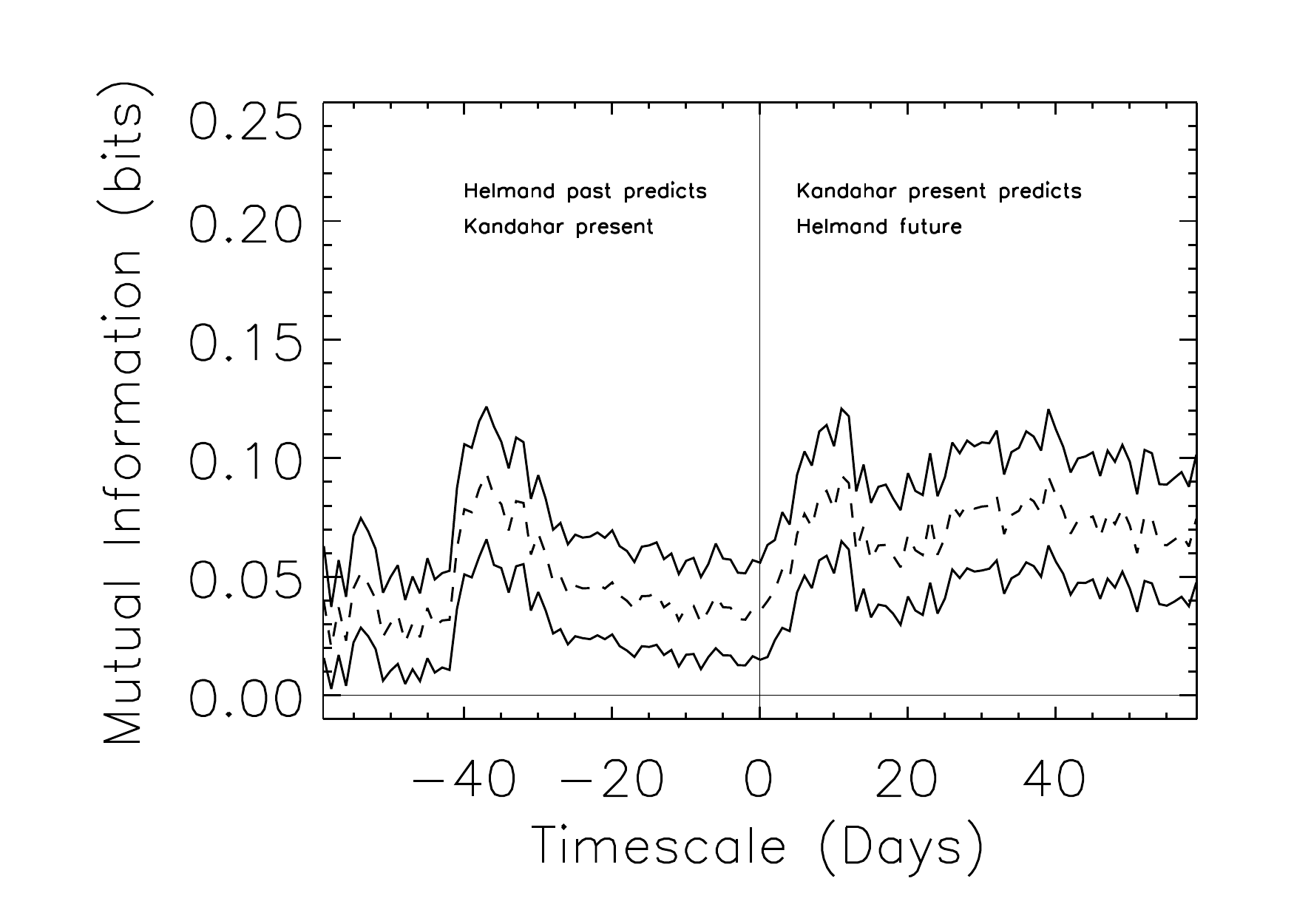}
\caption{Predictability of the Kandahar and Helmand time streams. Top: a dramatic asymmetry on short timescales provides strong suggestion of anticipatory, and potentially causal, effects transmitted \emph{from} Kandahar \emph{to} Helmand province on rapid (less than two-week) timescales. Bottom: the consistent, opposite asymmetry is seen in the reverse process. A rise in the predictability of Kandahar by Helmand on longer (one-month) timescales, mirrored in the top panel, suggests potentially longer-term seasonal or constraint-based information common to both systems.}
\label{h-k-compare}
\end{figure}

Figure~\ref{h-k-compare} shows how mutual information can be used to determine both the timescale and directionality of information flow. We consider the mutual information between a single day in one province, and the \emph{modal} day, on some date range, for the second province. Taking the modal day (\emph{i.e.}, the most common symbol found in that date range, with ties being broken by taking the median of the modes) is essentially a coarse-graining of the exponentially large multi-day state space; it scrambles time information within the range, and amounts (in the language of renormalization theory) to a particular decimation choice.

The top panel of Figure~\ref{h-k-compare} considers the mutual information between a day in Helmand province, and the modal day of either a range of dates in the past (negative $x$-axis) or future. For example, knowledge of the modal day for the prior twenty days in Kandahar leads to approximately 0.1 bits of information about the current day in Helmand. We can turn that phrasing around for the positive $x$-axis, and note that knowledge of current day's events in Helmand provide much \emph{less} information about the modal day in Kandahar's near future. 

This effect is reversed (lower panel) where we find that knowledge of Kandahar's present gives more information about Helmand's future modal days than the reverse. On longer timescales, there is some increase in the predictive power of Helmand's past for Kandahar's future ($-35$ days on the $x$-axis), which could potentially be attributed to knowledge of seasonal properties that affect both provinces---a similar predictive power is seen on the same timescales, in the same position, for the top panels, suggesting a common cause with similar effects.

\subsection{The Data Processing Inequality}
\label{dpi}

A novel relationship that arises for the case of mutual information is the \emph{data processing inequality}. In its simplest form, as found in ~\cite{cover}, it states that if three random variables $X$, $Y$ and $Z$ form a \emph{Markov Chain} in that order, \emph{i.e.},
\begin{equation}
p(x,y,z) = p(z|y)p(y|x)p(x)
\end{equation}
then
\begin{equation}
I(X,Y) \geq I(X,Z)
\label{dpi-eq}
\end{equation}
This is called the data processing inequality because the transformation from $Y$ to $Z$ can be seen as an (independent) ``processing'' of the output of a measurement of $Y$, which cannot (under the assumption of a perfectly rational observer) add any new information about $X$. Directly relevant to our work here is a four-term version of the inequality, where two underlying random variables, $X$ and $Y$, are known by two independent post-processings, $A$ and $B$. For us, the two mappings, $X\rightarrow A$ and $Y\rightarrow B$, are a good description of how a massively multi-dimensional random variable, describing the full state of a province, is reduced to the sum of the observed SIGACTs and, further, to the four-state codebook considered here. If the structure of the required conditional independencies is 
\begin{equation}
p(a,x,b,y) = p(x,y)p(a|x)p(b|y)
\end{equation}
then the mutual information between $A$ and $B$ is bounded by that between the underlying system,
\begin{equation}
I(X,Y) \geq I(A,B).
\label{tpi-eq}
\end{equation}
Equations~(\ref{dpi-eq}) and~(\ref{tpi-eq}) play a role similar to the Lin and Bhattacharyya bounds for decision-making in Section~\ref{b-bound-section}. If we are interested in the coordination of violence between provinces, then measurement of the mutual information between the dimensionality-reduced data provides a strict bound to the full system; phrased in the language of Section~\ref{b-bound-section}, it bounds the predictability, by a rational agent with at least as much information, from below.

A simpler form of the data processing inequality can be found for the case where the post-processing is deterministic, \emph{i.e.}, the entries of $p(a|x)$ and $p(b|y)$ are only either zero or one. This version of the data processing inequality amounts to the statement that coarse-graining a process will not, on average, increase the mutual information with a second data stream. This extends to any combination of deterministic coarse-graining and remapping (\emph{i.e.}, uniform reclassification of one of the event spaces).

\begin{eqnarray}
&& I\left(\left\{
\begin{array}{ccc}
p_{11} & p_{12} & p_{13} \\
p_{21} & p_{22} & p_{23} \\
\end{array} \right\}\right) \nonumber \\
&& = I\left(\left\{
\begin{array}{cc}
p_{11} & (p_{12}+p_{13}) \\
p_{21} & (p_{22}+p_{23}) \\
\end{array} \right\}\right) \nonumber \\
&& +\,(1-(p_{11}+\,p_{21}))I\left(\left\{
\begin{array}{cc}
p_{12} & p_{13} \\
p_{22} & p_{23} \\
\end{array} \right\}\right) \label{combine}
\end{eqnarray}
We study the preservation of this axiomatic relationship in detail in Section~\ref{cg-tests}.

\section{The Bootstrap Estimators In Practice}
\label{numerics}

The previous two sections have introduced two applications of information theory to the study of large-scale collective behavior. The framework allows us to quantify conceptually important aspects of both systems. Since estimation of these quantities is often from noisy, sample-limited data, we would like to know how our estimators perform in practice. We consider here the performance of the bootstrap estimators in the preservation of coarse-graining consistency, and in the error estimates and bias correction they provide. 

This section provides the main technical, as opposed to empirical or conceptual, results of our paper. Additional characterizations appear in Appendix~\ref{referee-performance}. The C and Python codes for the rapid evaluation of the Wolpert and Wolf, NSB and Bootstrap estimators used in this paper have been made publicly available as part of the THOTH package~\cite{thoth-url}.

\subsection{The Bayesian Prior Hierarchy}
\label{bph}

In order to characterize the bootstrap, we must first consider the range of problems we hope to apply it to. Doing this amounts to defining an ``underlying'' distribution over distributions. For the discrete-symbol processes we have considered above, a mathematically elegant choice for the underlying distribution involves the use of the Dirichlet distribution. For the homogenous case, the Dirichlet distribution is parameterized by a single parameter, $\beta$, where the probability of any particular distribution, $\vec{p}$, arising is
\begin{equation}
P(\vec{p}|D_\beta) = \frac{1}{Z(\beta)}\prod_{i=1}^k p_i^{\beta-1}
\label{uniform-d}
\end{equation}
where $Z(\beta)$ is a normalization constant. While $D_1$, sometimes called the ``Laplace prior'', is a common choice, it was noted by ~\cite{nemenman-additional,Nemenman:2004p18597} that it has unusual information-theoretic properties; in particular, the average value of the entropy of a distribution drawn from $D_1$ is quite high. Nemenman {\em et al.}~\cite{nemenman-additional,Nemenman:2004p18597} suggested an interesting alternative: to construct a \emph{mixture} of Dirichlet distributions $D_\mathrm{NSB}$ such that the entropy of a distribution drawn from $D_\mathrm{NSB}$ is approximately uniform.

The question of Bayesian estimators for information-theoretic quantities under a Dirichlet prior was addressed by~\cite{Wolpert:1995p17421} (WW), where tools were provided for estimation of entropy and mutual information for arbitrary $\beta$, and explicit formulas for the $D_1$ case were given; these are analytic in terms of polygamma functions. Nemenman {\em et al.}~\cite{Nemenman:2004p18597} provided a numerical method for the estimation of entropy under the $D_\mathrm{NSB}$ priors. In order to produce an NSB estimator for the mutual information, we can extend the Theorem 10 in ~\cite{Wolpert:1995p17421} to the $\beta\neq 1$ case, and then integrate this over the NSB prior (Equation~9 in~\cite{Nemenman:2004p18597}); this estimates mutual information under the assumption that draws of the $p_{ij}$ are (approximately) uniform in entropy.

In the Bayesian framework, more general choices of prior allow for the proper evaluation of a wider range of models; conversely, Bayesian estimators will often fail when given a sample whose underlying model lies outside the prior support. The $D_\mathrm{NSB}$ prior is more general than the $D_1$ prior, and as such has a wider range of applicability; we expect the NSB estimators to strongly outperform the WW estimators when evaluated on distributions drawn from Dirichlet distributions with $\beta$ much less than unity, for~example. 

Both priors, however, assume that bins are drawn from distributions with homogenous weights \linebreak (\emph{i.e.}, $\beta$ a constant independent of $i$ in Equation~(\ref{uniform-d})). This assumption is likely to fail in real world systems, and its failure may lead to inaccurate inferences. This is particularly problematic in biological and social systems where there are few clues to the correct choice of binning: if a three state system is best modeled as a draw from $D_\mathrm{NSB}$, then this assumption will fail for a different observer, who, under the influence of a rival theory, gathers data in such a way as to group two of the bins together to get a two-state system that now draws from an inhomogeneous distribution.

While a single Dirichlet distribution with inhomogeneous weights leads to asymmetries that may be hard to justify \emph{a priori}, this symmetry may be restored in Dirichlet mixtures. We thus consider in this paper a novel mixture, $D^\prime$, that allows a distribution to be drawn from a range of inhomogeneous Dirichlet distributions. A draw of a distribution $p$ with $k$ bins from $D^\prime$ is made as follows:
\vspace{-6pt}
\begin{enumerate}[1.]
\item Draw a random integer, $k^\prime$, between $k$ and $k^2$ inclusive. \label{size}
\vspace{-12pt}
\item Draw a distribution $p^\prime$ with $k^\prime$ bins from $D_\mathrm{NSB}$. Then $p^\prime$ is approximately uniform in entropy over $k^\prime$~bins.\label{nsb-draw}
\vspace{-12pt}
\item Randomly partition the $k^\prime$ bins into $k$ bins.\label{nsb-partition}
\vspace{-12pt}
\item Coarse grain the distribution $p^\prime$, given the partition of Step~\ref{nsb-partition}, to get $p$.
\vspace{-6pt}
\end{enumerate}
This construction always amounts to a draw from some Dirichlet distribution (since a coarse-graining of a Dirichlet distribution is itself a Dirichlet distribution with different parameters). Our use of random partitions restores bin symmetries and ensures that we are not placing unwarranted \emph{a priori} structures on the average properties of draws. Random partitioning may be done rapidly by modification of the {\tt ranksb} and {\tt rancom} algorithms from ~\cite{comb-book}.

Note that just as $D_1$ is strictly contained within $D_\mathrm{NSB}$, $D_\mathrm{NSB}$ is strictly contained within $D^\prime$ (in particular, $D_\mathrm{NSB}$ is selected when $k^\prime$ is drawn equal to $k$). This gives us a hierarchy of Bayesian priors,
\vspace{-6pt}
\begin{equation}
D_1 \subset D_\mathrm{NSB} \subset D^\prime
\vspace{-6pt}
\end{equation}
where the set containment here is interpreted in terms of the support of the distributions. 

Inhomogeneous Dirichlet mixtures of the $D^\prime$ form are extensions of the $D_\mathrm{NSB}$ prior. They are particularly useful in cases where coarse-graining plays a significant role: either because an observer has coarse-grained the system during the process of data-gathering, or because the system itself is expected to have provided only coarse-grained information about its underlying function. These conditions are particularly likely to obtain when studying social, cognitive, and biological systems. Our example of semantic coarse-graining, introduced in Section~\ref{cg}, where the number of bins used in the analysis ranges over more than two orders of magnitude provides an example. If one believes the $D_\mathrm{NSB}$ to be the optimal prior at a fine-grained level (say, the 26,740-dimensions at the word-stem level), a prior of $D^\prime$ form is thereby urged for higher, more coarse-grained, levels (such as the 116-dimensional binning used \linebreak in Section~\ref{distances}). 

We can oppose these cases, where coarse-graining is an essential and intrinsic part of the study, to the situation that obtains in many physical systems. There, the ``true'' binning may already be known from the underlying laws of physics, and so a different prior may be of greater use.

The $D^\prime$ has not been a focus of study in the context of Bayesian inference, and much work remains to be done. Step~\ref{size} clearly admits different parameterizations, for example, as do modifications of the partitioning algorithm for Step~\ref{nsb-partition}. We consider the question of extending Bayesian NSB methods to these inhomogeneous mixtures an interesting problem for future work. In this paper, our goal is to compare how well the Wolpert and Wolf and NSB estimators compare against non-Bayesian bootstrap methods. We answer this question in the following section.

\subsection{Coarse-Graining Consistency}
\label{cg-tests}

The coarse-graining consistency relationship for entropy, Condition~\ref{con1}, and the associated \linebreak coarse-graining relationship for mutual information, Equation~(\ref{combine}), are central to basic \linebreak information-theoretic results. These include the chain rules for entropy and mutual information (Theorems~2.5.1 and~2.5.2 in~\cite{cover}), which means that the standard Venn diagrams that dictate the relationships between entropies (joint and marginal) and mutual information hold.

The approximate satisfaction of Condition~\ref{emp2}$^\prime$ allows us to recover, approximately, nearly all of the structure of Information Theory in the finite-data limit. For example, since a conditional entropy, $H(X|Y)$, can be turned into a difference of entropies, $H(X,Y)-H(Y)$, and each term can be consistently estimated by $H_\mathrm{corr}$, our method allows us to estimate \emph{any} information-theoretic formula that analytically decomposes into the sum of entropies and conditional entropies, and bias-correct if so~desired.

For these reasons, we study how well the bootstrap preserves Condition~\ref{con1}. Because the satisfaction of Equation~(\ref{combine}) is directly related to the deterministic version of the data-processing inequality, we also test Equation~(\ref{combine}) directly. Our results follow directly from sampling $p$s from $D^\prime$, and then characterizing the performance of the different methods in estimating (known) properties of $p$ from finite samples drawn from $p$. We present these results in terms of the sampling factor, defined as the number of observations divided by the number of system states. 

Table~\ref{entropy-cg} shows that bootstrap estimator provides a large gain {\sc rms} consistency compared with two commonly-used estimation methods. Table~\ref{mi-cg} considers the mutual information consistency relation, Equation~(\ref{combine}). In a similar fashion to the case of Table~\ref{entropy-cg}, we find that the bootstrap has much improved performance compared with both WW and NSB. 

\begin{table}[H]
\centering
\begin{tabular}{cccc}
\toprule {\bf Sampling} & {\bf WW (RMS bits)} & {\bf NSB (RMS bits)} & {\bf Bootstrap (RMS bits)} \\ \midrule 1$\times$ & 0.2333 & 0.1376 & 0.0046 \\
2$\times$ & 0.1697 & 0.0865 & 0.0031 \\
4$\times$ & 0.1165 & 0.0463 & 0.0014 \\
8$\times$ & 0.0720 & 0.0239 & 0.0008 \\
16$\times$ & 0.0423 & 0.0121 & 0.0005 \\
\bottomrule
\end{tabular}
\caption{RMS Violations of coarse-graining consistency for entropy (Condition~\ref{con1}) for the Wolpert \& Wolf (WW), Nemenman, Shafee \& Bialek (NSB), and the bootstrap. The bootstrap estimator leads to a factor of ten or more improvement in coarse-graining consistency; as the amount of data increases, the bootstrap approaches full consistency faster. The average entropy of the three-state distributions is approximately 1.2 bits. These results are for the $D^\prime$ prior of Section~\ref{bph}.}
\label{entropy-cg}
\end{table}

\begin{table}[H]
\centering
\begin{tabular}{cccc}
\toprule {\bf Sampling} & {\bf WW (RMS bits)} & {\bf NSB-MI (RMS bits)} & {\bf Bootstrap (RMS bits)} \\ \midrule
1$\times$ & 0.0316 & 0.0387 & 0.0026 \\
2$\times$ & 0.0196 & 0.0323 & 0.0012 \\
4$\times$ & 0.0119 & 0.0208 & 0.0006 \\
8$\times$ & 0.0069 & 0.0124 & 0.0004 \\
16$\times$ & 0.0039 & 0.0066 & 0.0003 \\
\bottomrule
\end{tabular}
\caption{RMS Violations of coarse-graining consistency for mutual information \protect\linebreak (Equation~(\ref{combine})) for the Wolpert \& Wolf (WW) estimator, Nemenman, Shafee \& Bialek (NSB) for Mutual Information, and the bootstrap. The bootstrap estimator again leads to a factor of ten or more improvement in coarse-graining consistency; as the amount of data increases, the bootstrap approaches full consistency faster. The average mutual information of the $2\times3$ distribution is approximately 0.25 bits.}
\label{mi-cg}
\end{table}

The relatively poor performance of these two estimators is in part due to the homogenous nature of the $D_1$ and $D_\mathrm{NSB}$ priors, which place equal weight on all known bins. The true system (equal weight in three bins, lopsided weights in the coarse-grained version) is not contained within either space of priors.

In both cases, we consider coarse-grainings of distributions drawn from $D^\prime$; however, our results are largely insensitive to whether we use $D_1$ or $D_\mathrm{NSB}$ instead. In particular, the relative performance of the bootstrap and the Bayesian estimators is unchanged. The results of Tables~\ref{entropy-cg} and~\ref{mi-cg} are a main technical result of this paper.

The entropy consistency relationship of Condition~\ref{emp2} and Table~\ref{entropy-cg} is directly relevant to the analyses of Section~\ref{distances} of the predictability of trial outcomes at the Old Bailey. There we are concerned with a range of different possible ways to coarse-grain the underlying trial transcripts, all of which represent strong and contrasting theories of linguistic semantics, and none of which are given to us \emph{a priori}. Preservation of coarse-graining consistency means that we will not find anomalous gains or losses in predictive power by changing our underlying theories of social cognition and predictive abilities.

Meanwhile, the mutual information consistency relationship, Equation~(\ref{combine}) and Table~\ref{mi-cg}, is particularly useful in analysis of information flows between highly complex underlying state spaces, as was considered in the case of Afghanistan in Section~\ref{mi-big}~\cite{care}. Simplifying the codebook, or enlarging it to account for additional features of the SIGACTs or other parallel data streams, will, again, lead to consistent shifts in the estimated levels of coordination, signaling and predictability that reflect the structure of the underlying system, and not features of the prior space.

\subsection{Bias Correction and the Reliability of Error Estimates}
\label{bias-tests}

Having established the utility of the bootstrap in preserving information-theoretic axioms, we conclude this technical section by characterizing the reliability of its bias correction and error estimates. We ask, in other words, how well we estimate the quantities in question, and how well we estimate our uncertainty about them. In this section, we neglect cases where the empirical distribution has entropy zero; these cases form a separate class of problem for which the bootstrap is particularly unsuited.

Our use of the bootstrap involves a bias correction, and so we first want to know how well the correction works and how that correction compares to other methods in the literature. For a particular $\vec{p}$, bias is defined as
\begin{equation}
B_H(n, \vec{p}) = \langle H_\mathrm{est}(\vec{n}) \rangle_{P(\vec{n}|\vec{p})} - H(\vec{p})
\end{equation}
for the case of entropy, and similarly for mutual information and Jensen--Shannon divergence. Informally, $B_H(n, \vec{p})$ asks what the average difference is between an estimate of $H$, from a sample of $n$ observations drawn from $\vec{p}$, and the true value, $H(\vec{p})$. In many cases, unbiased estimators are possible; as noted in Section~\ref{cg}, however, estimates of information theoretic quantities are necessarily biased and the real question is the extent to which this bias is reduced by appropriate choice of estimator.

The bias will vary depending on the particular $\vec{p}$ chosen; for simplicity, we consider the average bias for probabilities that are themselves drawn from a distribution; explicitly, we consider
\begin{equation}
B_H(n, D) = \langle B_H(n, \vec{p}) \rangle_{P(\vec{p}|D)}
\end{equation}
The use of a Bayesian estimator with prior $D$ guarantees that $B_H(n,D)$ will be zero (note that this does not violate the results of ~\cite{Paninski:2003ff}---bias for any particular $\vec{p}$ may be non-zero); as discussed above, since we draw from an inhomogeneous prior, $D^\prime$, that is strictly larger than either of the priors used in our Bayesian estimators, this study also allows us to characterize the performance of NSB and WW when predicting out of class.

We also want to know how trustworthy our error estimates are. One useful way to quantify this is to ask how often the true value of the quantity in question lies within the $1\sigma$ (one standard deviation, or 68.2\%) and $2\sigma$ (two standard deviations, or 95.4\%) ranges.

\begin{figure}[H]
\centering
\includegraphics[width=4.6in]{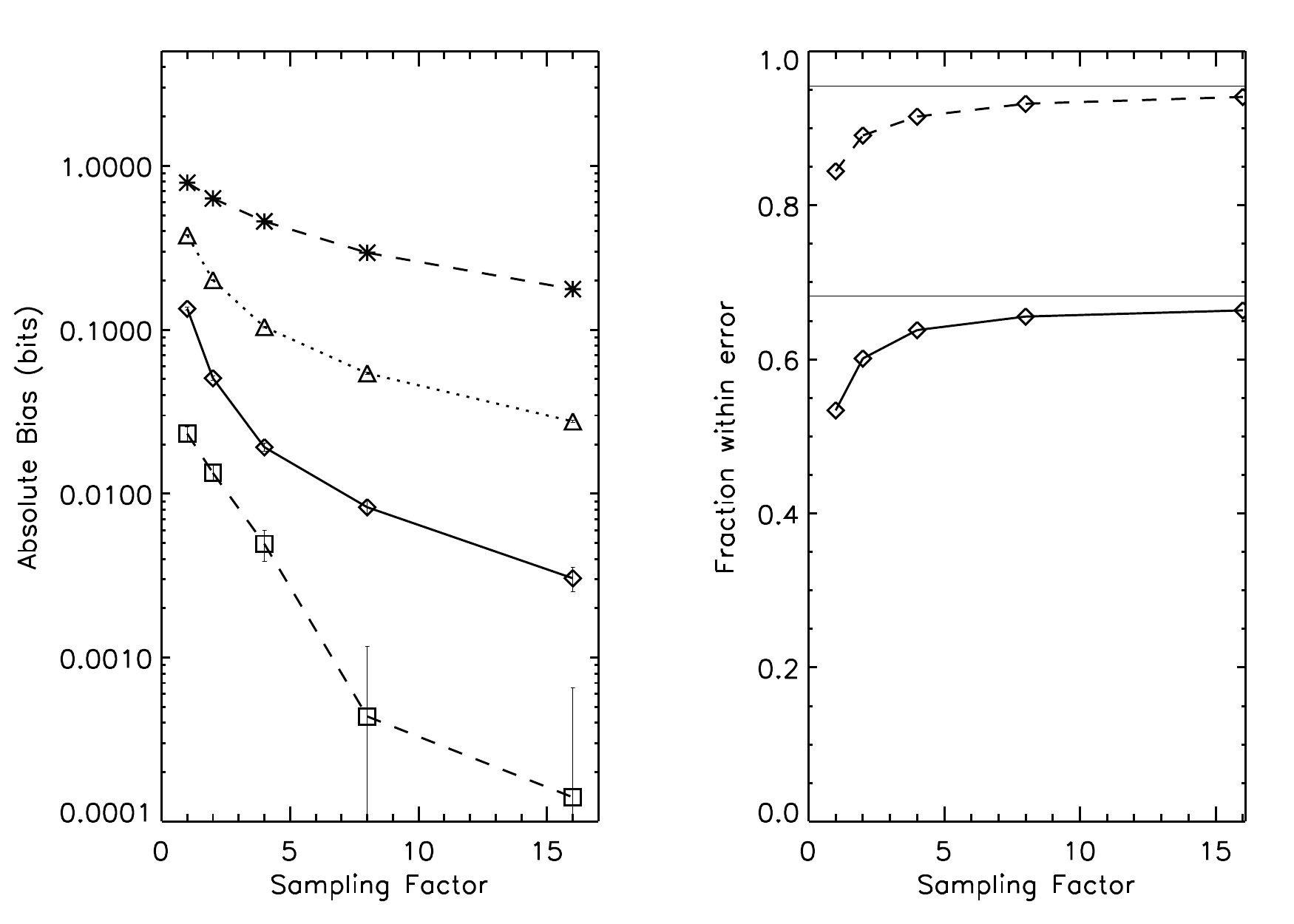}
\caption{Left panel: bias for the 16-state entropy estimation case, under prior $D^\prime$. Dotted line: naive estimator; Dashed line, *-symbol: Wolpert and Wolf estimator; Dashed line, $\square$-symbol: NSB estimator. Solid line: Bootstrap estimator. Right panel: one-sigma (solid line) and two-sigma (dashed line) error bar reliability; as the sampling factor increases, both rapidly approach their asymptotic values (thin horizontal lines). Average entropy for this prior is 2.4 bits.}
\label{16ent}
\end{figure}

\begin{figure}[H]
\centering
\includegraphics[width=4.5in]{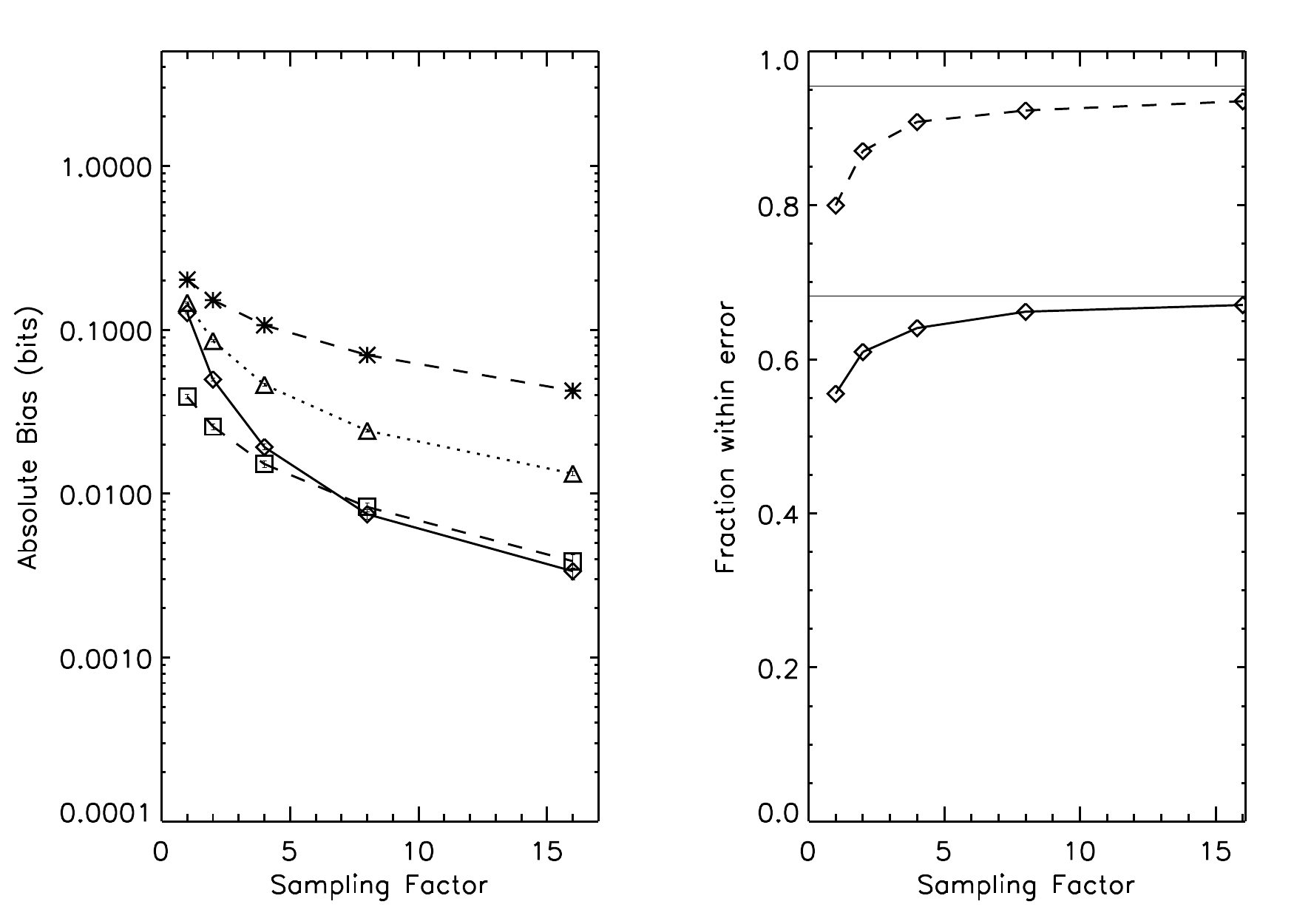}
\caption{Left panel: bias for the estimation of mutual information on a $4\times4$ joint probability distribution, under prior $D^\prime$. Dotted line: naive estimator; Dashed line, *-symbol: Wolpert and Wolf estimator; Dashed line, $\square$-symbol: NSB estimator. Solid line: Bootstrap estimator. Right panel: one-sigma (solid line) and two-sigma (dashed line) error bar reliability; as the sampling factor increases, both rapidly approach their asymptotic values (thin horizontal lines). Average mutual information under this prior is 0.55 bits.}
\label{4mi}
\end{figure}

\begin{figure}[H]
\centering
\includegraphics[width=5in]{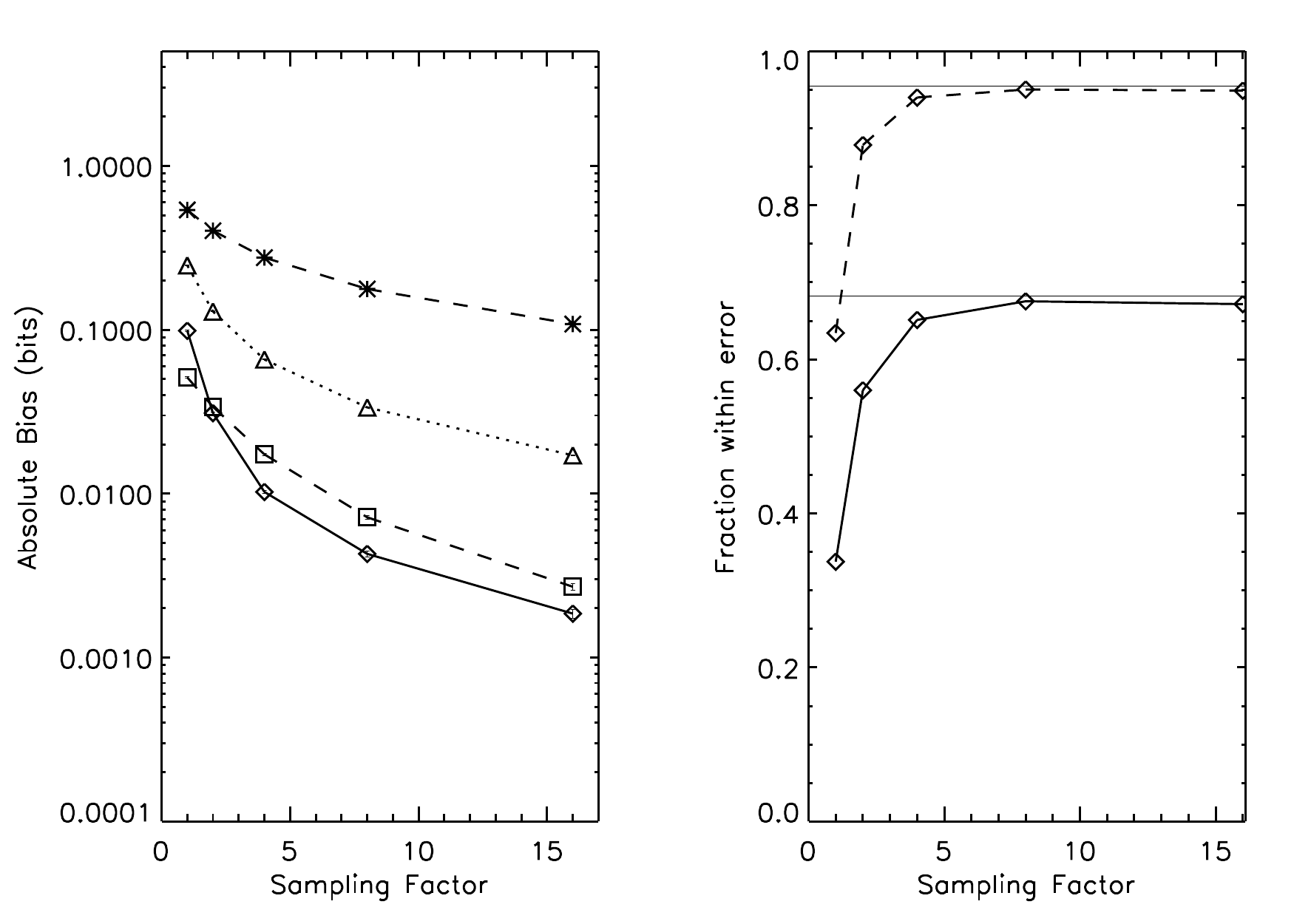}
\caption{Left panel: bias for the estimation of mutual information on a $16\times16$ joint probability distribution, under prior $D^\prime$. Dotted line: naive estimator; Dashed line, *-symbol: Wolpert and Wolf estimator; Dashed line, $\square$-symbol: NSB estimator. Solid line: Bootstrap estimator. Right panel: one-sigma (solid line) and two-sigma (dashed line) error bar reliability; as the sampling factor increases, both rapidly approach their asymptotic values (thin horizontal lines). Average mutual information for this prior is 1.33 bits.}
\label{16mi}
\end{figure}

Our results on bias and error reliability are shown for two test cases---the estimation of the entropy of a 16-state system (Figure~\ref{16ent}), and the estimation of mutual information for a $4\times 4$ joint probability (Figure~\ref{4mi}). Figure~\ref{16mi} shows the $16\times16$ case.
\vspace{-12pt}
\vspace{-12pt}

Bias correction often comes at a cost, since the relationship between the true bias and its estimated value is itself noisy. We find, however, that the {\sc rms} error of the bootstrap estimator is comparable to our best Bayesian estimator, NSB; this is shown in Tables~\ref{ent-rms}--\ref{mi-rms-16}. 

\begin{table}[H]
\centering
\begin{tabular}{cccc}
\toprule {\bf Sampling} & {\bf WW (RMS bits)} & {\bf NSB (RMS bits)} & {\bf Bootstrap (RMS bits)} \\ \midrule
1$\times$ & 1.0442 & 0.3305 & 0.3605 \\
2$\times$ & 0.8302 & 0.2223 & 0.2304 \\
4$\times$ & 0.5990 & 0.1537 & 0.1564 \\
8$\times$ & 0.3854 & 0.1039 & 0.1051 \\
16$\times$ & 0.2341 & 0.0730 & 0.0734 \\
\bottomrule
\end{tabular}
\caption{RMS errors for estimation of the entropy information of a $16$-state system, for the Wolpert \& Wolf (WW) estimator, Nemenman, Shafee \& Bialek (NSB) for Mutual Information, and the bootstrap. The bootstrap estimator has {\sc rms} errors comparable to the~NSB.}
\label{ent-rms}
\end{table}

\begin{table}[H]
\centering
\begin{tabular}{cccc}
\toprule {\bf Sampling} & {\bf WW (RMS bits)} & {\bf NSB (RMS bits)} & {\bf Bootstrap (RMS bits)} \\ \midrule
1$\times$ & 0.3106 & 0.1924 & 0.2783 \\
2$\times$ & 0.2494 & 0.1446 & 0.1722 \\
4$\times$ & 0.1851 & 0.1041 & 0.1132 \\
8$\times$ & 0.1277 & 0.0727 & 0.0761 \\
16$\times$ & 0.0845 & 0.0523 & 0.0534 \\
\bottomrule
\end{tabular}
\caption{RMS errors for estimation of the mutual information of a $4\times4$ joint probability, for the Wolpert \& Wolf (WW) estimator, Nemenman, Shafee \& Bialek (NSB) for Mutual Information, and the bootstrap. The bootstrap estimator has {\sc rms} errors comparable \protect\linebreak to the NSB.}
\label{mi-rms}
\end{table}
\begin{table}[H]
\centering
\begin{tabular}{cccc}
\toprule {\bf Sampling} & {\bf WW (RMS bits)} & {\bf NSB (RMS bits)} & {\bf Bootstrap (RMS bits)} \\ \midrule
1$\times$ & 0.6920 & 0.1028 & 0.1417 \\
2$\times$ & 0.5267 & 0.0685 & 0.0642 \\
4$\times$ & 0.3696 & 0.0417 & 0.0373 \\
8$\times$ & 0.2409 & 0.0260 & 0.0251 \\
16$\times$ & 0.1487 & 0.0175 & 0.0173 \\
\bottomrule
\end{tabular}
\caption{RMS errors for estimation of the mutual information of a $16\times16$ joint probability, for the Wolpert \& Wolf (WW) estimator, Nemenman, Shafee \& Bialek (NSB) for Mutual Information, and the bootstrap. The bootstrap estimator has {\sc rms} errors comparable \protect\linebreak to the NSB.}
\label{mi-rms-16}
\end{table}

\subsection{Summary}

Our results in this section provide strong support for the use of the bootstrap in the parameter ranges of relevance to many estimation problems. While the NSB estimator is unbiased over the $D^\prime$ space (within error) for the entropy estimation case, it does violate the coarse-graining consistency relationship much more strongly. Meanwhile, the bootstrap is comparable to the NSB in bias and {\sc rms} error when used for estimation of more sophisticated quantities such as mutual information. 

As an example of the use of the characterizations of this section, Figure~\ref{4mi} allows us to read off, directly, useful information necessary for the evaluation of the claims of Section~\ref{mi-big} for asymmetries of information flow in Afghanistan. For the $4\times4$ mutual information estimation problem, we have at least 305 days worthy of observations (for the edges of the $\pm 60$ day range). This amounts to a $19\times$ oversampling, putting us on the right-hand edge of each panel in Figure~\ref{4mi}. Our bias is well below the overall signal, while our estimation of the $1\sigma$ error band is seen to be reliable.

\newpage

\section{Conclusions}
\label{conc}

We have presented two case studies of the use of information theory for the scientific study of collective phenomena associated with cognitively complex social systems. Scientifically compelling accounts of the role of information---such as those that involve reference to optimal prediction or signaling---often rely on some of the central axioms of information theory, including coarse-graining consistency, error-rate bounds and the data processing inequality. In studying the nature of our (current) tool of choice, then, we have characterized the (approximate) preservation of these axioms in the use of these tools in ranges likely to be relevant to current data.

Information theory originated in the need to describe, and place limits, on the ability of engineered systems to communicate and process signals, to infer properties of the outside world, and to tolerate risk and uncertain environments. The extension of information-theoretic concepts from engineered systems and inferential tasks to the biological and social sciences expands the domain of the theory and places the measurement of its quantities at center stage. 

In many cases, the role of human, animal, or otherwise evolved reason in a natural system means that information theory is just as relevant there as it is for the study of engineered systems. Improvements in our understanding of both biological and social systems in large part depend upon increasing our understanding of how they encode and process information. Many systems devote significant amounts of constrained resources to precisely these tasks: the representation of aspects of the environment, and transformations on those representations---as opposed to direct intervention in the environment itself. 

The design principles for evolved systems may be very different, but the underlying laws are the same. Just as the study of an organism's structural morphology will make reference to engineering concepts such as efficiency, stability and dissipation, so will accounts of how individuals behave in ambiguously cooperative environments make reference to information-theoretic concepts such as bounds on \linebreak optimal predictions.

Less obviously, but no less central, is the role of coarse-graining in the construction of scientific accounts of collective phenomena. 

In the physical sciences, coarse-graining is a fundamental part of the construction of theories. In condensed matter and quantum field theory, the notion of a renormalization group is based on spatial proximity: things that are physically near each other can be grouped together, and the theory relating the properties of these coarse-grained groups can be related to the theory corresponding to the finer-grained description in a systematic fashion. 

In the case of biological and social systems, functional and computational principles dominate over physical proximity, and the parallel construction is in its infancy~\cite{DeDeo:2011p19975}. In the interim, we often find it necessary to conduct analyses of such systems using informally-derived coarse-grainings dictated by a combination of domain-specific intuition (which properties should be grouped together), numerical or analytic tractability (often, how much grouping is needed for reliable statistics, rejection of null hypotheses, and so forth), and the richness and accuracy of the underlying data itself. Indeed, in cognitive systems ranging from the neurobiological to the social, the emergence of this coarse-graining is itself a pressing scientific question~\cite{Olshausen:1997cr,Olshausen:2004fw,Daniels:2012vs}.

For both these reasons---the desire to study the role of reasoning in nature and our lack of knowledge about the right way to carve nature at its joints~\cite{plato}, it is useful not only (on the one hand) to estimate information-theoretic quantities but also (on the other) to derive functions of the data that obey its underlying axioms. The technical characterization of the bootstrap in Section~\ref{numerics} provides strong support for its use in place of other estimators when these axioms become important to the reasoning one wants to do. Future work in this field will almost certainly provide better tools---including, one hopes, a fully Bayesian method for preserving the axioms exactly---and new insights into the role that reason and inference play in the natural world.

\section*{Acknowledgements}
\vspace{12pt}

SD thanks John Geanakoplos, Cosma Shalizi, Jim Crutchfield, Ryan James and David Wolpert for helpful conversations, and acknowledges the support of a Santa Fe Institute (SFI) Omidyar Postdoctoral Fellowship. An early version of this work was presented at ``Combing Information and Game Theory," 15 August 2012, supported by SFI, the New Mexico Consortium, and Bill and Stephanie Sick. RH thanks Drew Cabaniss for helpful conversations, and acknowledges the support of an REU at SFI under National Science Foundation (NSF) Grant SMA-1005075. SK acknowledges the support of an SFI Graduate Fellowship. SD, RH and SK acknowledge the support of NSF Grant EF-1137929, ``The Small Number Limit of Biological Information Processing,'' and of the Emergent Institutions project.

\section*{Conflict of Interest} 
\vspace{12pt}

The authors declare no conflict of interest.

\appendix
\section*{Appendix}
\vspace{-12pt}
\section{The Trial of John Long, as Reported on 18 September 1820}
\label{jlong}

{\sc John Long} was indicted for that he, on the 28th of August, upon George North, feloniously, wilfully, and maliciously did make an assault, and with a sharp instrument did strike and cut him, in and upon his face, with intent to kill and murder him, and do him some grievous bodily harm.

{\sc Ann Hickman}. I live in Gardner's lane, King street, Westminster. The prisoner and his wife lived in the same house. On the 28th of August I was looking out of window, and saw the prisoner and his wife in the yard, with North, consulting about parting---his wife had cohabited with North, and wanted her clothes to go away with him; the prisoner said if she came up stairs she should have them---then all three came into the passage. I heard words at the foot of the stairs, went down, and saw North against the wall---he said, ``I am done for." I saw his hand drop against the wall, it was all bloody. The prisoner was about half a yard from him. I did not see him do anything, and never heard him threaten North. I saw a knife in his hand, and said ``Long take care what you are at, and give me the knife.'' He shut the knife up, and gave it to me---it was rather bloody. North had three cuts in his face; I saw nothing more. He was taken to the doctor's. A quantity of blood laid at the door. Whatever happened was done before I came down. 

{\sc Henry Betts}. I am a constable. About eight o'clock on the night this happened, I was sent for, and knocked at the prisoner's room door, and told him to open it. I found it open, he was there in bed with his wife, she was in liquor. I took him to the watch-house. Four or five days after I saw North, his wounds were dressed. He had one cut from his ear down towards his mouth, his lip was cut, and he had a stab in his cheek. 

Prisoner's Defence. My wife had deserted me, and gone with North; I met them together, and North said if I touched her he would break every bone in my body. She followed me home, I was going to take her up stairs, and he seized me by the throat.

{\sc George North}, being called, did not appear. 

[Total length: 398 words. Verdict: not guilty. Old Bailey Online Trial ID t18200918-112]

\section{SIGACT for IED Explosion, 3 April 2005, Kandahar Province}
\label{sigact}

{\bf Title}

IED ANP CIV Other 1 CIV KIA 1 ANP KIA 

{\bf Text}

CJSOTF REPORTS IED STIKE IVO SPIN BULDAK. THE FOLLOWING SALT REPORT WAS SENT: S: 1X IED, A: IED STRKE, L: 42RTV 51120 32190, T: 0315Z. REMARKS: FR SOF SENDING OUT RECON TO INVESTIGATE. AT 0722Z TG ARES CURRENTLY ON SITE CONDUCTING INVESTIGATION. THERE WAS ONE EXPLOSION FROM A MINE HIDDEN IN A BEVERAGE VENDORS CART. APPEARS TO HAVE BEEN COMMAND DETONATED. 1X DISTRICT POLICE OFFICER IS KIA. NO WIA REPORTED.

{\bf Additional Data}

{\tt \small \{:reportkey=>"DCEAC77F-A84D-45F3-88B3-33B9B5A95B20", :type=>:explosivehazard, :category=>:iedexplosion, :trackingnumber=>:2007-033
-005423-0737, :region=>:rcsouth, :attackon=>:enemy, :complexattack=>false, :reportingunit=>:other, :unitname=>:other, :typeofunit=>:noneselected, :friendlywia=>0, :friendlykia=>0, :hostnationwia=>0, :hostnationkia=>1, :civilianwia=>0, :civiliankia=>1, :enemywia=>0, :enemykia=>0, :enemydetained=>0, :mgrs=>"42RTV5110332180", :latitude=>30.99694061, :longitude=>66.39333344, :originatorgroup=>:unknown, :updatedbygroup=>:unknown, :ccir=>:"", :sigact=>:"", :affiliation=>:enemy, :dcolor=>:red, :classification=>:secret, :start=>2005-04-03 03:15:00 UTC, :province=>:kandahar, :district=>:spinboldak, :nearestgeocode=>"AF241131834", :nearestname=>"Spin Boldak"\}}

{\bf Source and Post-processing}

Original source Ref.~\cite{wl-set}; post-processing for initiative (via methods of ~\cite{larry-article}), geocode (GPS cross-reference to data provided by Afghanistan Information Management Service), and additional filtering by collaboration~\cite{hawkins}.

\section{Additional Characterizations}
\label{referee-performance}

The expectation values that form the definition of the bootstrap have a natural expression in terms of sums over the multinomial distribution,
\begin{equation}
\langle H(\vec{n}^*) \rangle_{P(\vec{n}^* | \vec{n})}=\sum_{\vec{n}^* \in\mathbb{N}^*} \frac{n!}{n^*_1!\cdots n^*_k!}\left(\frac{n_1}{n}\right)^{n^*_1}\cdots\left(\frac{n_k}{n}\right)^{n^*_k}
\label{bootstrap-explicit}
\end{equation}
where $\mathbb{N}^*$ is the subspace of $k$-dimensional vectors with $L^1$ norm equal to $n$. The bootstrap is thus non-linear in $\vec{n}$; its bias correction is of a different nature from estimators linear, or piecewise-linear, in $\vec{n}$, such as the Miller--Madow~\cite{miller}. Direct calculation of Equation~(\ref{bootstrap-explicit}) is exponentially hard; as with all bootstrap estimators of which we are aware, the most efficient method is to approximate of $H_\mathrm{corr}(\vec{n})$ by Monte Carlo re-sampling.

The bias correction presented in this paper can be thought of as a zeroth-order approximation. In particular, it assumes that the bias of the resampled data is equal to the bias of the actual data (in Figure~\ref{bootstrap}, that the two red lines are, on average for that particular underlying distribution, identical). Because the naive entropy under-estimates the true entropy, and because this bias increases monotonically with entropy, this zeroth-order correction will always reduce the bias. This is strictly true only if one allows use of the bootstrap on empirical distributions with entropy zero, a usually rare case as noted above; since our analysis neglects these cases, Figure~\ref{entropy-function} shows a slightly positive bias at the smallest entropies.

More sophisticated corrections allow for a functional dependence of bias on the underlying estimator. In many cases, the benefit to linear or non-linear corrections can be significant~\cite{MacKinnon:1998bx,MacKinnon:2006bs}. These \linebreak higher-order corrections, however, do not decompose in the simple fashion of Equation~(\ref{dir1}) and Equation~(\ref{dir2}), and will violate the coarse-graining consistency relations much more strongly.

The bootstrap, naive, and NSB estimators perform differently as a function of the entropy of the underlying distribution. This is shown in Figure~\ref{entropy-function}, where we measure the average bias as a function of the entropy of the underlying distribution, when we draw from $D^\prime$. The NSB estimator (dashed line) tends to overestimate low-entropy samples, while the naive (dotted) and bootstrap (solid) tend to underestimate in the same regime. The results for $D_\mathrm{NSB}$ are similar, though (of course) the overall average bias for the NSB estimator is now zero.

The various prior distributions examined in this paper, $D_\beta$, $D_{\mathrm{NSB}}$ and $D^\prime$, lead to different distributions over entropy. Figure~\ref{entropy-dist} shows the distribution of entropies for probability distributions drawn from $D_{1}$, $D_{\mathrm{NSB}}$ and $D^\prime$, for the case of draws from a sixteen-category distribution. As can be seen, the $D_{\mathrm{NSB}}$ distribution leads to an impressively flat distribution of entropies above one bit (further numerical work suggests that transition to flatness at one bit does not change as the size of the space increases). The $D^\prime$ distribution shifts entropies somewhat to the center of the range. The $D_{1}$ distribution, otherwise known as the Laplace Prior, has the overwhelming majority of its prior support near the maximum possible entropy, making it hard to accumulate evidence for low-entropy distributions.

\setcounter{figure}{0}
\renewcommand\thefigure{A\arabic{figure}}

\begin{figure}[H]
\centering
\includegraphics[width=4in]{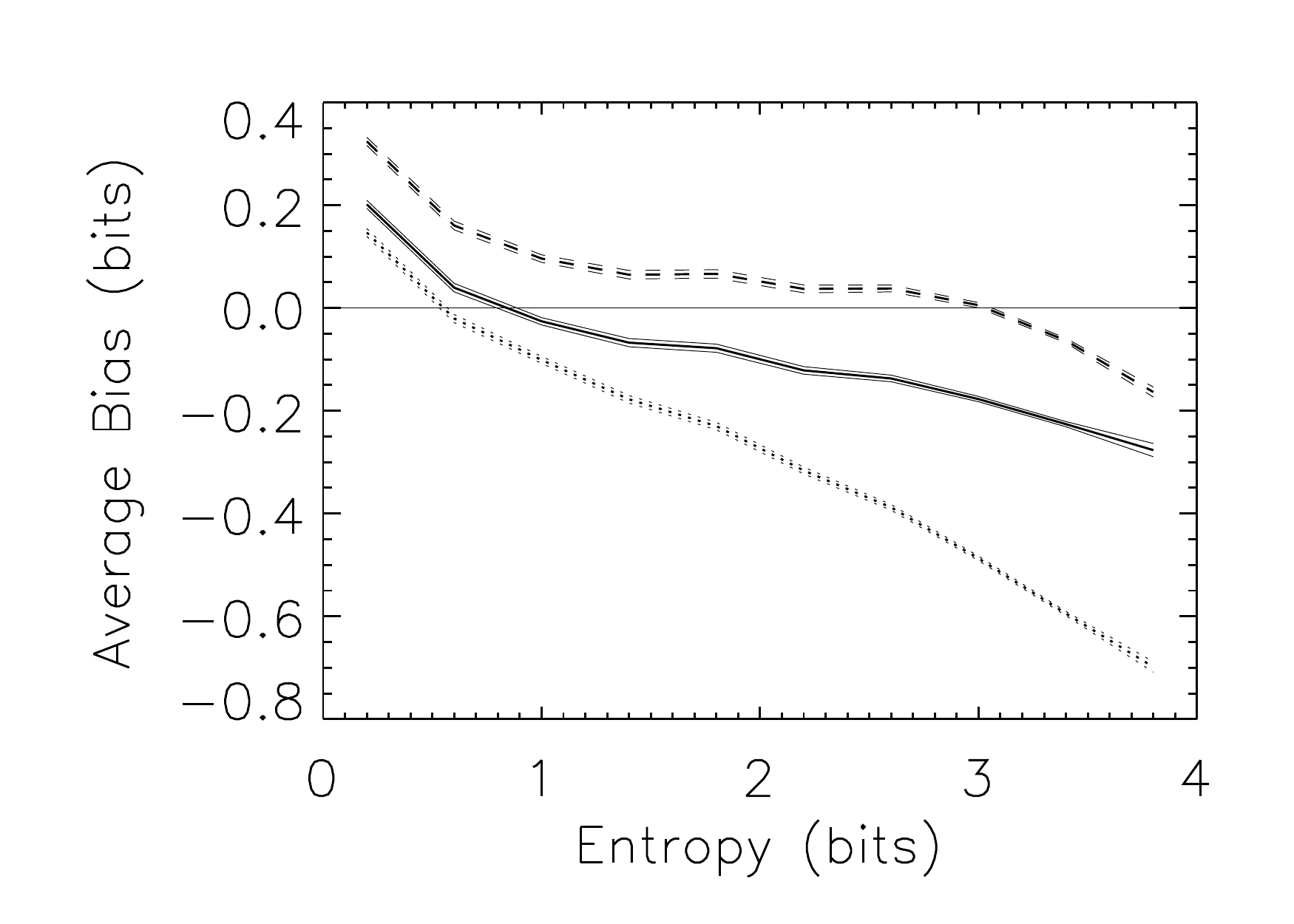}
\caption{Estimator bias as a function of the entropy of the underlying distribution for the naive (dotted line), NSB (dashed line) and bootstrap (solid line) estimators. Distributions are over sixteen categories, drawn from $D^\prime$, and binned in 0.25 bit increments; the bias is for estimates made with sixteen samples (\emph{i.e.}, $1\times$ oversampling). Ranges shown are one-sigma error bars for the bias in the bin. As can be seen, all estimators tend to overestimate small entropies, and underestimate large entropies, with the cross-over point (and overall bias) depending on the method. As in the main text, Section~\ref{bias-tests}, we neglect cases where the empirical distribution has entropy zero; this is one source of the positive bias at the lowest entropy bins.}
\label{entropy-function}
\end{figure}

\begin{figure}[H]
\centering
\includegraphics[width=4in]{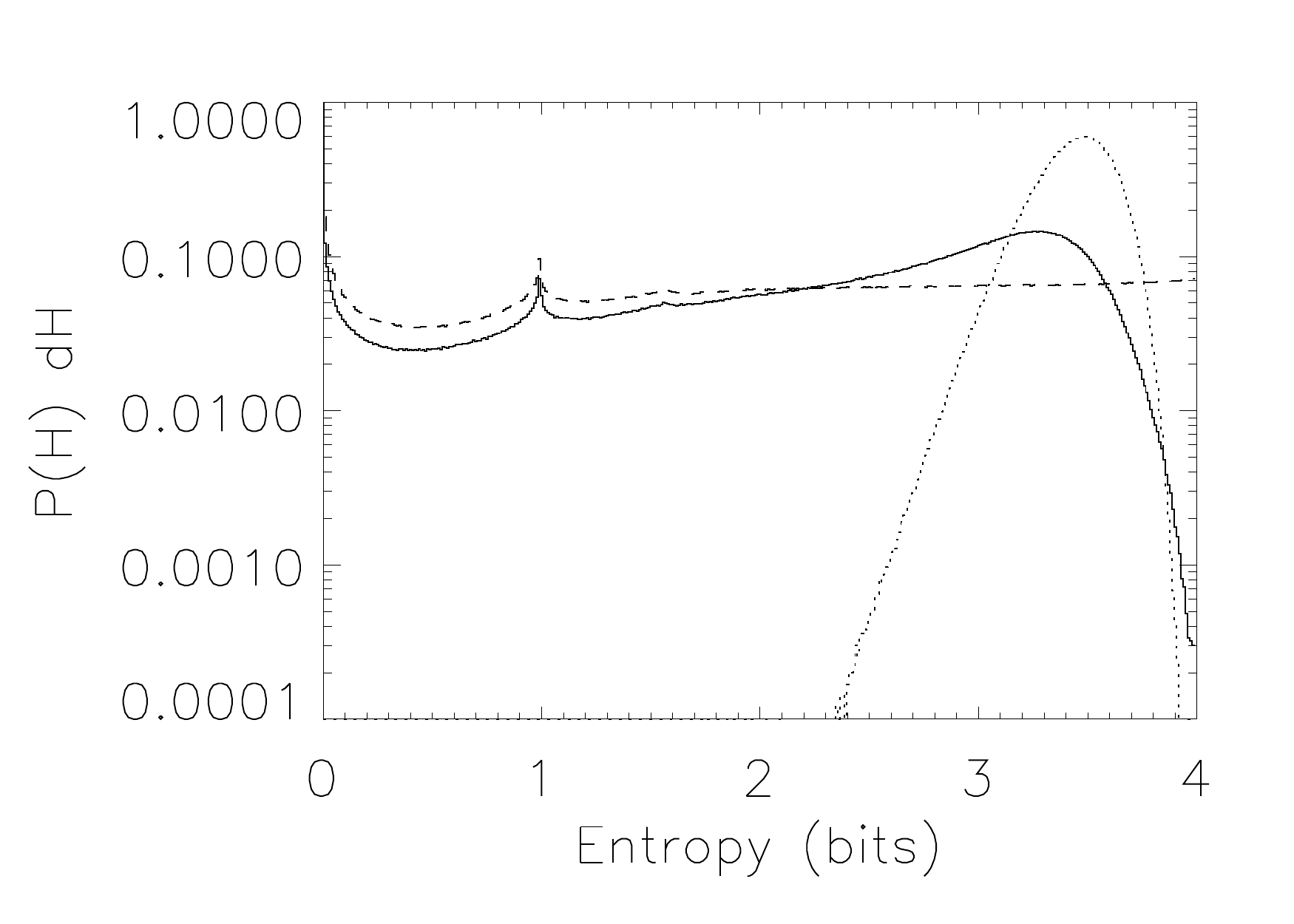}
\caption{The distribution of entropies for distributions sampled from the $D_1$ (dotted line), $D_\mathrm{NSB}$ (dashed line) and $D^\prime$ (solid line) priors. The $D_1$ prior produces distributions very strongly skewed towards high entropies, while the $D_\mathrm{NSB}$ distribution is nearly flat for entropies larger than one bit.}
\label{entropy-dist}
\end{figure}

\bibliographystyle{mdpi}

\begin{thebibliography}{-------}
\providecommand{\natexlab}[1]{#1}

\bibitem[Shannon(1948)]{Shannon:1948p18105}
Shannon, C.
\newblock {A mathematical theory of communication}.
\newblock {\em Bell System Technical Journal} {\bf 1948}, {\em 27},~379--423,
  623--656.

\bibitem[Paninski(2003)]{Paninski:2003ff}
Paninski, L.
\newblock {Estimation of Entropy and Mutual Information}.
\newblock {\em Neural Computation} {\bf 2003}, {\em 15},~1191--1253.

\bibitem[sar()]{sara}
Collaboration ``Institution Formation, Semantic Phenomena, and Three Hundred
  Years of the British Criminal Court System'', with participants Simon DeDeo,
  Sara Klingenstein, and Tim Hitchcock.

\bibitem[wor()]{word-order}
This is not exactly true, since we use part-of-speech information for our
  semantic classifier, and the means (a hidden Markov model) for identifying a
  word's lexical category are sensitive to order and context: compare ``the
  dogs bit the sailor'' [`dogs' as plural noun] and ``the sailor dogs the
  waitress'' [`dogs' as verb]. It is an interesting and open question the
  extent to which syntactical features convey semantic information.

\bibitem[mag()]{magic-word}
It is possible to correct the magic word problem \emph{post hoc}, merging
  categories until no magic words remain. If this rule is specified with
  sufficient generality, it allows for bootstrap estimation. We do not consider
  this approach here.

\bibitem[Lin(1991)]{Lin:1991ug}
Lin, J.
\newblock {Divergence measures based on the Shannon entropy}.
\newblock {\em IEEE Transactions on Information Theory} {\bf 1991}, {\em
  37},~145--151.

\bibitem[Nielsen(2010)]{Nielsen:2010vk}
Nielsen, F.
\newblock {A family of statistical symmetric divergences based on Jensen's
  inequality}.
\newblock {\em eprint arXiv:1009.4004} {\bf 2010}, {\em cs.CV}.

\bibitem[Hellman and Raviv(1970)]{Hellman:1970fn}
Hellman, M.; Raviv, J.
\newblock {Probability of error, equivocation, and the Chernoff bound}.
\newblock {\em IEEE Transactions on Information Theory} {\bf 1970}, {\em
  16},~368--372.

\bibitem[Jaynes and Bretthorst(2003)]{jaynes}
Jaynes, E.; Bretthorst, G.
\newblock {\em Probability Theory: The Logic of Science}; Cambridge University
  Press,  2003.

\bibitem[Cover and Thomas(2006)]{cover}
Cover, T.; Thomas, J.
\newblock {\em Elements of Information Theory}; Wiley: Hoboken, New Jersey,
  USA,  2006.

\bibitem[Chen(1976)]{Chen:1976jm}
Chen, C.H.
\newblock {On information and distance measures, error bounds, and feature
  selection}.
\newblock {\em Information Sciences} {\bf 1976}, {\em 10},~159--173.

\bibitem[Ito(1972)]{ito1972approximate}
Ito, T.
\newblock Approximate error bounds in pattern recognition.
\newblock {\em Machine Intelligence} {\bf 1972}, {\em 7},~369--376.

\bibitem[Hashlamoun \em{et~al.}(1994)Hashlamoun, Varshney, and
  Samarasooriya]{Hashlamoun:1994gp}
Hashlamoun, W.A.; Varshney, P.K.; Samarasooriya, V.N.S.
\newblock {A tight upper bound on the Bayesian probability of error}.
\newblock {\em IEEE Transactions on Information Theory} {\bf 1994}, {\em
  16},~220--224.

\bibitem[bs-()]{bs-jsd}
We can estimate $\rho$ itself by means of the statistical bootstrap. Resampling
  suggests that $\rho$ itself may suffer from bias which can be corrected for.

\bibitem[fur()]{further-clarify}
Consider, for example, the case where the supports of $p_1$ and $p_2$ only
  partially overlap: $p_1(x)$ is zero for some $x$, but $p_2(x)$ is non-zero,
  and vice versa for some $y$. The draw $\{x,y\}$ is not possible for either
  $p_1$ or $p_2$, but is possible for the average.

\bibitem[O'Loughlin \em{et~al.}(2010)O'Loughlin, Witmer, Linke, and
  Thorwardson]{larry-article}
O'Loughlin, J.; Witmer, F.; Linke, A.; Thorwardson, N.
\newblock Peering into the Fog of War: The Geography of the WikiLeaks
  Afghanistan War Logs 2004--2009.
\newblock {\em Eurasian Geography and Economics} {\bf 2010}, {\em 51},~472--95.

\bibitem[Zammit-Mangion \em{et~al.}(2012)Zammit-Mangion, Dewar,
  Kadirkamanathan, and Sanguinetti]{2012PNAS..10912414Z}
Zammit-Mangion, A.; Dewar, M.; Kadirkamanathan, V.; Sanguinetti, G.
\newblock {Point process modelling of the Afghan War Diary}.
\newblock {\em Proceedings of the National Academy of Sciences} {\bf 2012},
  {\em 109},~12414--12419.

\bibitem[ana()]{analysis-detail}
There are more than 78,000 SIGACTs in the original data set. We removed
  approximately 20,000 because they expressed a level of ambiguity about what
  happened (``unknown-initiated action'' or ``suspicious incident''), because
  their GPS coordinates did not match the description, because they were not
  time-sensitive (``weapons cache found'') or because they were irrelevant to
  our study (``IED hoax'' or ``show of force''). Information on initiative is
  explicitly included in the SIGACT data and is extracted by the methods of
  Ref.~\cite{larry-article}.

\bibitem[San{\'\i}n and Giustozzi(2010)]{Sanin:2010cx}
San{\'\i}n, F.G.; Giustozzi, A.
\newblock {Networks and Armies: Structuring Rebellion in Colombia and
  Afghanistan}.
\newblock {\em Studies in Conflict {\&} Terrorism} {\bf 2010}, {\em
  33},~836--853.

\bibitem[Gutierrez~Sanin(2008)]{GutierrezSanin:2008el}
Gutierrez~Sanin, F.
\newblock {Telling the Difference: Guerrillas and Paramilitaries in the
  Colombian War}.
\newblock {\em Politics {\&} Society} {\bf 2008}, {\em 36},~3--34.

\bibitem[Green(2011)]{green}
Green, A.H.
\newblock Repertoires of Violence Against Non-combatants: The Role of Armed
  Group Institutions and Ideologies.
\newblock PhD thesis, Yale University,  2011.

\bibitem[haw()]{hawkins}
Collaboration ``The Emergence of Insurgency in Afghanistan: an Information
  Theoretic Analysis'', with participants Simon DeDeo and Robert Hawkins.

\bibitem[{Hutter}(2001)]{Hutter:2001wt}
{Hutter}, M.
\newblock {Distribution of Mutual Information}.
\newblock {\em eprint arXiv:cs/0112019} {\bf 2001}.
\newblock also as Advances in Neural Information Processing Systems 14
  (NIPS-2001) pg. 399-406.

\bibitem[Zaffalon and Hutter(2002)]{Zaffalon:2002vx}
Zaffalon, M.; Hutter, M.
\newblock {Robust feature selection by mutual information distributions}.
\newblock {\em Proceedings of the Eighteenth Conference on Uncertainty in
  Artificial Intelligence} {\bf 2002}, pp. 577--584.

\bibitem[{Williams} and {Beer}(2011)]{2011arXiv1102.1507W}
{Williams}, P.L.; {Beer}, R.D.
\newblock {Generalized Measures of Information Transfer}.
\newblock {\em eprint arXiv:1102.1507} {\bf 2011}, {\em physics.data-an}.

\bibitem[Schreiber(2000)]{2000PhRvL..85..461S}
Schreiber, T.
\newblock {Measuring Information Transfer}.
\newblock {\em Physical Review Letters} {\bf 2000}, {\em 85},~461--464.

\bibitem[tho()]{thoth-url}
Available at \url{http://thoth-python.org}; last accessed 22 May 2013.

\bibitem[Nemenman \em{et~al.}(2002)Nemenman, Shafee, and
  Bialek]{nemenman-additional}
Nemenman, I.; Shafee, F.; Bialek, W.
\newblock {\em Advances in Neural Information Processing Systems} {\bf 2002},
  {\em 14}.

\bibitem[Nemenman \em{et~al.}(2004)Nemenman, Bialek, and de~Ruyter~van
  Steveninck]{Nemenman:2004p18597}
Nemenman, I.; Bialek, W.; de~Ruyter~van Steveninck, R.
\newblock {Entropy and information in neural spike trains: Progress on the
  sampling problem}.
\newblock {\em Physical Review E} {\bf 2004}, {\em 69},~56111.

\bibitem[Wolpert and Wolf(1995)]{Wolpert:1995p17421}
Wolpert, D.H.; Wolf, D.R.
\newblock {Estimating functions of probability distributions from a finite set
  of samples}.
\newblock {\em Physical Review E} {\bf 1995}, {\em 52},~6841.

\bibitem[{Nijenhuis} and {Wilf}(1978)]{comb-book}
{Nijenhuis}, A.; {Wilf}, H.S.
\newblock {\em {Combinatorial algorithms for computers and calculators}};
  Computer Science and Applied Mathematics, Academic Press,  1978.
\newblock 2nd ed.

\bibitem[car()]{care}
Care needs to be taken, however, since a general stochastic function of the
  underlying process will not, in general, preserve independence in the
  empirical distribution. It is always possible, for example, to choose an
  instantiation of a stochastic re-mapping \emph{post hoc} that magnifies
  accidental correlations found in the original empirical distribution. We
  thank John Geanakoplos for pointing this out to us.

\bibitem[DeDeo(2011)]{DeDeo:2011p19975}
DeDeo, S.
\newblock {Effective theories for circuits and automata}.
\newblock {\em Chaos} {\bf 2011}, {\em 21},~7106.

\bibitem[Olshausen and Field(1997)]{Olshausen:1997cr}
Olshausen, B.A.; Field, D.J.
\newblock {Sparse coding with an overcomplete basis set: A strategy employed by
  V1?}
\newblock {\em Vision Research} {\bf 1997}, {\em 37},~3311--3325.

\bibitem[Olshausen and Field(2004)]{Olshausen:2004fw}
Olshausen, B.A.; Field, D.J.
\newblock {Sparse coding of sensory inputs}.
\newblock {\em Current opinion in neurobiology} {\bf 2004}, {\em 14},~481--487.

\bibitem[Daniels \em{et~al.}(2012)Daniels, Krakauer, and Flack]{Daniels:2012vs}
Daniels, B.C.; Krakauer, D.C.; Flack, J.C.
\newblock {Sparse code of conflict in a primate society}.
\newblock {\em Proceedings of the National Academy of Sciences} {\bf 2012},
  {\em 109},~14259--14264.

\bibitem[Plato(1925)]{plato}
Plato.
\newblock {\em Phaedrus}; Harvard University Press,  1925.
\newblock Plato in Twelve Volumes, Vol. 9. Translated by Harold N. Fowler; see
  \url{http://www.perseus.tufts.edu/hopper/text?doc=Plat.+Phaedrus+265e}, last
  accessed 22 May 2013.

\bibitem[wl-()]{wl-set}
Afghan War Diary.
\newblock Last accessed 4 June 2013 at
  \url{http://wikileaks.org/afg/event/2005/04/AFG20050403n68.html}.

\bibitem[Miller(1955)]{miller}
Miller, G.
\newblock Note on the bias of information estimates. In {\em Information theory
  in psychology II-B}; Quastler, H., Ed.; Free Press: Glencoe, IL,  1955.

\bibitem[MacKinnon and Smith(1998)]{MacKinnon:1998bx}
MacKinnon, J.G.; Smith, Jr., A.A.
\newblock {Approximate bias correction in econometrics}.
\newblock {\em Journal of Econometrics} {\bf 1998}, {\em 85},~205--230.

\bibitem[MacKinnon(2006)]{MacKinnon:2006bs}
MacKinnon, J.G.
\newblock {Bootstrap Methods in Econometrics}.
\newblock {\em Economic Record} {\bf 2006}, {\em 82},~S2--S18.

\end{thebibliography}

\end{document}